\documentclass[twocolumn,superscriptaddress,floatfix,preprintnumbers,prl,nofootinbib]{revtex4-2}
\usepackage[colorlinks=true,breaklinks=true]{hyperref}
\usepackage[utf8]{inputenc}
\usepackage{float}
\hypersetup{allcolors=[rgb]{0.0 0.0 0.6},linkcolor=[rgb]{0.75 0.05 0.05}}
\usepackage{amsmath,amssymb}
\usepackage{mathtools}
\usepackage[normalem]{ulem}
\usepackage{bm}
\usepackage{physics}
\usepackage{siunitx}
\DeclareSIUnit \parsec {pc}
\DeclareSIUnit \years {yr}
\DeclareSIUnit \electronvolt {eV}
\usepackage[dvipsnames]{xcolor}
\usepackage{float}
\usepackage{graphicx}
\usepackage{epsfig}
\usepackage{letltxmacro}
\LetLtxMacro{\oldcite}{\cite}
\renewcommand{\cite}[1]{\mbox{\oldcite{#1}}}

\usepackage{longtable,booktabs} 

\usepackage{orcidlink}

\long\def\exclude#1{}

\DeclareSymbolFont{starfontsym}{OT1}{sts}{m}{n}
\DeclareMathSymbol{\mathTerra}{\mathord}{starfontsym}{76}

\newcommand{\beq}{\begin{equation}}
\newcommand{\eeq}{\end{equation}}

\def\ga{\,\,\raise0.14em\hbox{$>$}\kern-0.76em\lower0.28em\hbox
{$\sim$}\,\,}

\newcommand{\e}{\mathrm{e}}
\newcommand{\iu}{\mathrm{i}}

\long\def\exclude#1{}

\allowdisplaybreaks

\bibpunct{[}{]}{,}{n}{}{,}

\hypersetup{
    colorlinks=true,
    citecolor=[rgb]{.1, .7, .6},
    linkcolor=[rgb]{.2, .55, .95},
    filecolor=magenta,
    urlcolor=[rgb]{.1, .7, .6},
}

\usepackage{appendix}

\begin{document}

\title{Ultra-Light Dark Matter Simulations and Stellar Dynamics: Tension in Dwarf Galaxies for $m < 5\times10^{-21} $ eV}

\author{Luca Teodori}
\affiliation{Department of Particle Physics and Astrophysics, Weizmann Institute of Science, Rehovot 7610001, Israel}
 \affiliation{Instituto de Astrof\'isica de Canarias, C/ V\'ia L\'actea, s/n E38205, La Laguna, Tenerife, Spain
 	} 
 \affiliation{Universidad de La Laguna, Departamento de Astrof\'isica, La Laguna, Tenerife,
 	Spain}

\author{Andrea Caputo} 
\affiliation{Department of Theoretical Physics, CERN, Esplanade des Particules 1, P.O. Box 1211, Geneva 23, Switzerland}
\affiliation{Dipartimento di Fisica, ``Sapienza'' Universit\`a di Roma \& Sezione INFN Roma1, Piazzale Aldo Moro
5, 00185, Roma, Italy}

\author{Kfir Blum}
\affiliation{Department of Particle Physics and Astrophysics, Weizmann Institute of Science, Rehovot 7610001, Israel}


\begin{abstract}
We present numerical simulations of dark matter and stellar dynamics in ultra light dark matter halos tailored to mimic dwarf galaxies. An important effect we observe is the dynamical evolution of the stellar half-light radius and velocity dispersion, which makes previous equilibrium models significantly incomplete. 
Based on half-light radius dynamical evolution, as well as velocity peaks due to soliton core condensation, we show that data from the Fornax, Carina, and Leo II dwarf galaxies disfavores particle masses in the range $ 5\times 10^{-22} \text{~eV} \lesssim m \lesssim 5\times10^{-21}$~eV. Smaller boson masses, around $m\approx1\times10^{-22}$~eV, could cause strong dynamical heating, but we caution that tidal stripping by the Milky Way could moderate the effect.
A caveat in our analysis is the omission of stellar self-gravity, which could affect extrapolation back in time if the stellar body was much more compact in the past.

\end{abstract}

\maketitle

\section{Introduction}. The fundamental nature of Dark Matter (DM) is a key open question in physics. As of the time of writing, only bounds and null hints have been placed on the properties of DM. This applies also to the most basic property: the mass of DM particles $m$. In this work we seek to advance the theoretical understanding of, and the resulting observational constraints on, the dynamics of DM near the lower limit on $m$. 

UltraLight Dark Matter (ULDM)~\cite{Hui2017,Hui:2021tkt,Ferreira:2020fam} is a scenario in which DM particles are so light, and thereby so numerous, that their dynamics is best described in terms of classical wave mechanics. 
ULDM fields can arise, for example, in models of string theory compactification, and the required cosmological relic abundance can be explained via vacuum misalignment~\cite{Svrcek:2006yi,Arvanitaki:2010sy,Hui2017}. ULDM was originally proposed to address specific shortcomings of the massive particle Cold Dark Matter (CDM) paradigm on small scales~\cite{Hu:2000ke}, but the motivation to study it is more general as this is, in effect, an attempt to establish the ultimate lower bound on $m$. 
This hinges on the fact that ULDM exhibits potentially observable signatures in various astrophysical systems through gravitation alone, without assuming any direct interaction between DM and the Standard Model.

On scales comparable to the de-Broglie wavelength of ULDM with velocity dispersion $\sigma$,
\begin{equation}
\frac{\lambda_{\rm dB}}{2\pi} \approx \frac{1}{m\sigma} \approx \SI{2}{\kilo\parsec} \qty(\frac{\SI{1e-22}{\electronvolt}}{m}) \qty(\frac{\SI{10}{\kilo\meter\per\second}}{\sigma} ) \ ,
\end{equation}
the wave nature of ULDM is manifest. In cosmology, wave-mechanical pressure suppresses structure formation, leading to constraints from the Cosmic Microwave Background (CMB) anisotropies and galaxy clustering~\cite{Lague:2021frh}, as well as from the Lyman-$\alpha$ forest, disfavoring (roughly) $m\lesssim10^{-21}$~eV~\cite{Irsic:2017yje,Armengaud:2017nkf,Kobayashi:2017jcf,Leong:2018opi}\footnote{A more ambitious bound was reported in Ref.~\cite{Rogers:2020ltq}. Keeping in mind systematic uncertainties potentially affecting Lyman-$\alpha$ analyses at the smallest scales (see, e.g. discussion in~\cite{Hui2017}), we believe that this bound requires further verification.}. More recently, Ref.~\cite{Nadler:2025fcv} showed that ULDM can strongly suppress the abundance of haloes below \(\lesssim 10^{9}\,M_\odot\), and derived a stringent lower bound, \(m \gtrsim 10^{-20}\,\mathrm{eV}\). The classical dwarf spheroidals studied in the present work, with stellar masses of order \(\sim 10^{7}\,M_\odot\), are expected to reside in substantially more massive haloes, with typical pre-infall (peak) masses \(\gtrsim 10^{10}\,M_\odot\) for Fornax and \(\gtrsim 10^{9}\,M_\odot\) for Leo~II, depending on the modelling assumptions~\cite{Fattahi_2018, Munshi_2021}. Our study is therefore complementary to halo mass-function constraints, which are subject to different systematics. For example, typical analyses implement ULDM effects only through modifications to the linear matter power spectrum in the initial conditions, and may therefore miss important non-linear effects that could affect these bounds. There have been attempts to perform cosmological simulations of FDM that fully model the dynamical effects of the quantum potential throughout cosmic evolution, such as Refs.~\cite{Elgamal:2023yzt, May:2022gus}, but these are still far from demonstrating sensitivity to \(m \gtrsim 10^{-22}\text{--}10^{-21}\,\mathrm{eV}\) at adequate resolution.

In galaxies, wave mechanics causes density interference patterns, and the formation of cored profiles (``solitons" in the literature) near the center of potential wells. These features are seen in simulations by many groups~\cite{Guzman2004,Schive:2014dra,Schive:2014hza,Schwabe:2016rze,Veltmaat:2016rxo,Mocz:2017wlg,Veltmaat:2018dfz,Eggemeier:2019jsu,Chen:2020cef,Schwabe:2020eac,Zhang:2018ghp, Marsh:2018zyw, Chiang:2021uvt} and are theoretically understood to a large extent~\cite{Chavanis:2011zi,Chavanis:2011zm,Levkov:2018kau,Guth:2014hsa,Bar-Or:2018pxz,Bar-Or:2020tys,Chan:2022bkz,Chan:2023crj,Chavanis:2019faf,Chavanis:2020upb}. The fluctuating density field is expected to affect dynamical heating, friction, and relaxation of the stellar body~\cite{Amorisco:2018dcn,Church:2018sro,Lancaster:2019mde,Bar:2021jff,Marsh:2018zyw, Dalal:2022rmp, Schive:2019rrw, Yang:2024hvb, DuttaChowdhury:2023qxg} (see also Ref.~\cite{Hertzberg:2022vhk} for an ULDM quantum tunnelling discussion relevant for dwarf galaxies). In the soliton region, quasinormal mode oscillations~\cite{Guzman2004,Chan:2023crj} and soliton random walk~\cite{Li:2020ryg} also contribute to heating\footnote{Additional purely gravitational probes of ULDM include~\cite{Khmelnitsky:2013lxt,Schutz:2020jox,Laroche:2022pjm,Powell:2023jns,Blum:2021oxj,Blum:2024igb, Blas:2024duy}.}.

We thus have two main differences of  ULDM with respect to CDM, as far as galaxy dynamics is concerned: a cored profile on scales of $\lambda_{\rm dB}$ in contrast with a cusp~\cite{Navarro:1996gj}, and dynamical heating from the soliton and the interference patterns. Given these features, it is important to understand whether ULDM halos are compatible with observations. Analyses of rotationally-supported low-surface-brightness galaxies led to the bound $m\gtrsim1\times10^{-21}$~eV~\cite{Bar:2018acw,Bar:2019bqz,Bar:2021kti} (see also~\cite{Bernal_2017}). These analyses, however, focused on the soliton signature alone, and relied on ULDM-only simulation results from Refs.~\cite{Schive:2014dra,Schive:2014hza} without self-consistent treatment of stellar dynamics. Ref.~\cite{Schive:2014dra}, in turn (see also \cite{Safarzadeh:2019sre, Gonzalez-Morales:2016yaf, Deng:2018jjz, Lora:2011yc}), focused on dispersion-dominated dwarfs, but they, too, did not simulate both the ULDM and the stars. 

Here we focus on dispersion-dominated dwarfs, and perform numerical simulations that include both the ULDM and the stars, the latter treated as test particles. 
We explore different initial conditions and ULDM particle masses in the range $\SI{1e-22}{\electronvolt} \le m/\si{\electronvolt} \le \SI{1e-20}{\electronvolt}$. Our simulations are tailored to resemble the Fornax, Leo II, and Carina dwarf galaxies, and we compare our results with available data~\cite{Walker:2008ax,Walker:2009zp,Koch:2007ye}.

The ULDM density profile in our simulations is initialized without a soliton core, but forms it dynamically in agreement with the soliton-halo relation discovered in Ref.~\cite{Schive:2014hza}. We show that the stellar line-of-sight velocity dispersion (LOSVD) approximately follows the steady solution to the collisionless Jeans' equation. However, dynamical heating leads to secular growth of the stellar half-light radius, with a time scale that decreases with decreasing $m$, qualitatively consistent with analytical expectations~\cite{Bar-Or:2018pxz,Bar-Or:2020tys}. Together, these effects put ULDM in tension with Fornax data for $m \lesssim \SI{1e-21}{\electronvolt}$. In particular, for $m=\SI{1e-22}{\electronvolt}$, the stellar body is contained inside the soliton region and LOSVD can in principle be reconciled with data; however, the dynamical heating timescale is much smaller than the system age of order $\SI{10}{\giga\years}$. This makes the half-light radius observed today difficult to explain without fine-tuning the initial conditions.  For $m=\SI{1e-21}{\electronvolt}$, dynamical heating becomes less important in Fornax; however, the soliton occupies only the inner part of the stellar body, causing a bump in LOSVD, in tension with the data.  The Leo II and Carina galaxies have smaller characteristic scales and smaller velocity dispersion than Fornax. Consequently, the analysis of these systems probes larger $m$: this can be attributed to the scaling properties of the Schr\"odinger-Poisson equation (SPE), that governs ULDM dynamics~\cite{Hui2017}; see App.~\ref{s:code}. Performing simulations for these systems we find that the tension with data, dominantly due to dynamical heating of the stellar population, extends up to $m \approx \SI{5e-21}{\electronvolt}$.

We notice that tidal stripping can significantly reduce dynamical heating for $m \simeq 10^{-22}$~eV, see Ref.~\cite{Yang:2025bae}. In fact, all Milky Way satellites experience some degree of tidal mass loss. However, the impact of tides on ULDM-specific signatures is not captured simply by noting that a system is a satellite. In the case of ULDM the relevant question is whether tidal interactions remove a substantial fraction of the outer halo/excited-state structure that sources the time-dependent ``granular'' potential fluctuations responsible for dynamical heating (see Ref.~\cite{Eberhardt:2025lbx}, the discussion in Sec.~III of Ref.~\cite{May:2025ppj} and our App.~\ref{s:tidal}). A useful diagnostic is therefore the separation of scales between the ULDM coherence length (set by the local de~Broglie wavelength, or equivalently the soliton/core scale) and the tidal radius at pericentre. For Fornax, this separation of scale is indeed small for the masses of interest, and tidal interactions are an important caveat for $m \simeq 10^{-22}$~eV. However, for other galaxies, such as Leo~II which we also study, the tidal radius remains much larger than the ULDM coherence scale for the ULDM masses of interest. Further details about this point will be provided in an upcoming publication~\cite{CaputoTeodori2026WhenTidesAreSmall}.

\begin{figure*}[ht]
    \centering
    \includegraphics[width=0.49\linewidth]{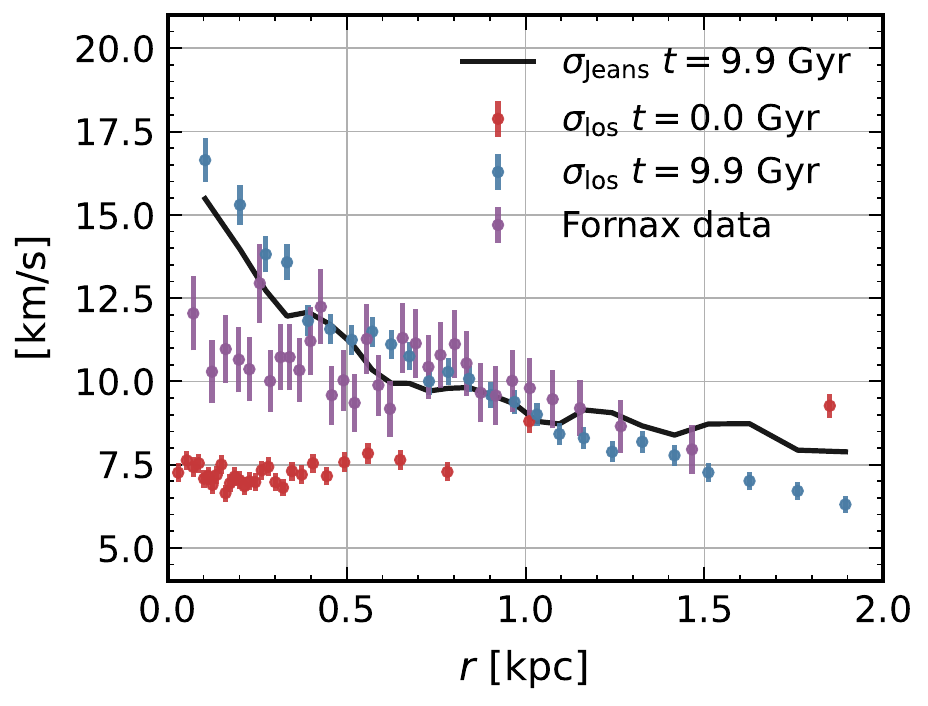}
    \includegraphics[width=0.475\linewidth]{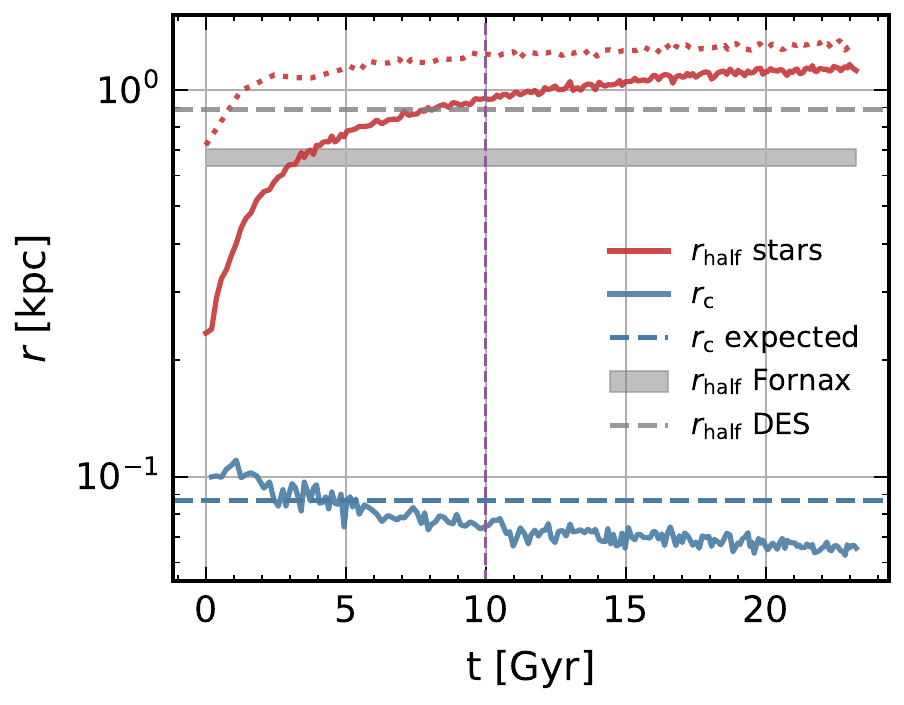}
    \caption{Simulation results for $m=\SI{1e-21}{\electronvolt}$, with $L=12$~kpc. 
    \textbf{Left.} LOSVD data for Fornax (purple) is compared to the simulation at time 0 (red) and at time $ t\approx\SI{9.9}{\giga\years} $ (blue).  Every bin has the same number of stars. Black line is the $\sigma_{\rm los}$ predicted by a Jeans analysis. Notice that LOSVD data are difficult to reconcile with ULDM for $r\lesssim0.5$~kpc. \textbf{Right.} Red lines: half-light radius $r_{\rm half}$ evolution over time. Solid and dotted refer to two simulations, with the same ULDM halo but different initial distribution of stars. The stellar kinematics analysis of the left plot is done for the lowest initial $r_{\rm half}$ simulation. Grey band indicates the observed half-light radius of Fornax reported in Tab.~I of Ref.~\cite{2009ApJ...704.1274W}, $r_{\rm half} = 0.668 \pm 0.034 \,\mathrm{kpc}$, while grey dashed line shows the half-light radius from Ref.~\cite{DES:2018jtu} (see discussion in App.~\ref{s:data}). 
    Blue line shows the soliton core radius $r_{\rm c}$. Dashed blue line shows the  expectation from the soliton-halo relation of Ref.~\cite{Schive:2014hza}. Vertical dashed line highlights $t=10$~Gyr, which is the typical age of the systems we consider. Notice that fine-tuned initial conditions allow to reproduce the half-light radius from~\cite{DES:2018jtu} (albeit not from~\cite{2009ApJ...704.1274W}).
    }
    \label{fig:NFW_1em21_main}
\end{figure*}

\begin{figure*}[ht]
    \centering
      \includegraphics[width=0.49\linewidth]{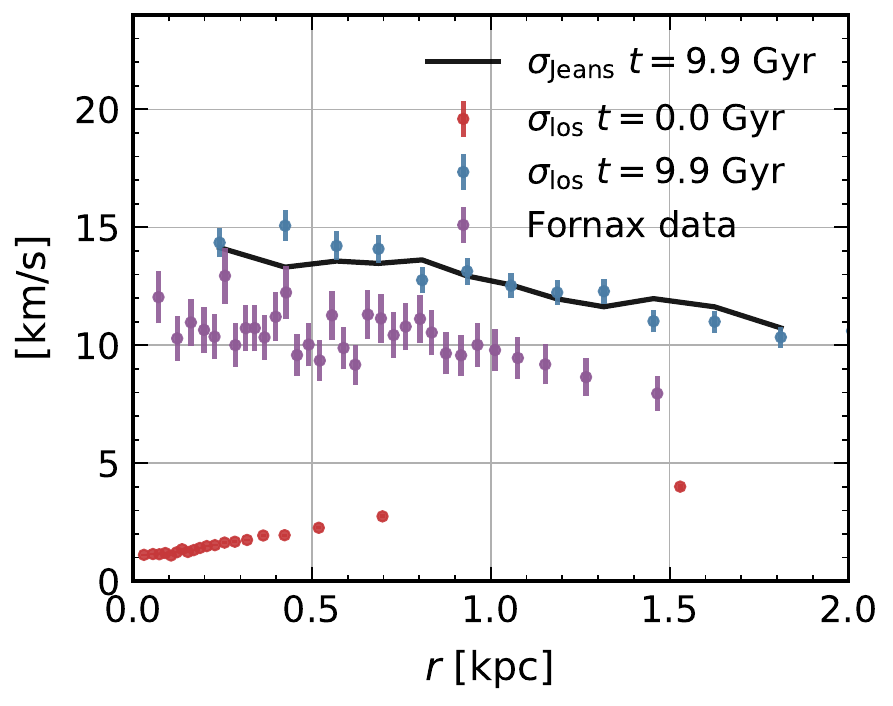}
      \includegraphics[width=0.49\linewidth]{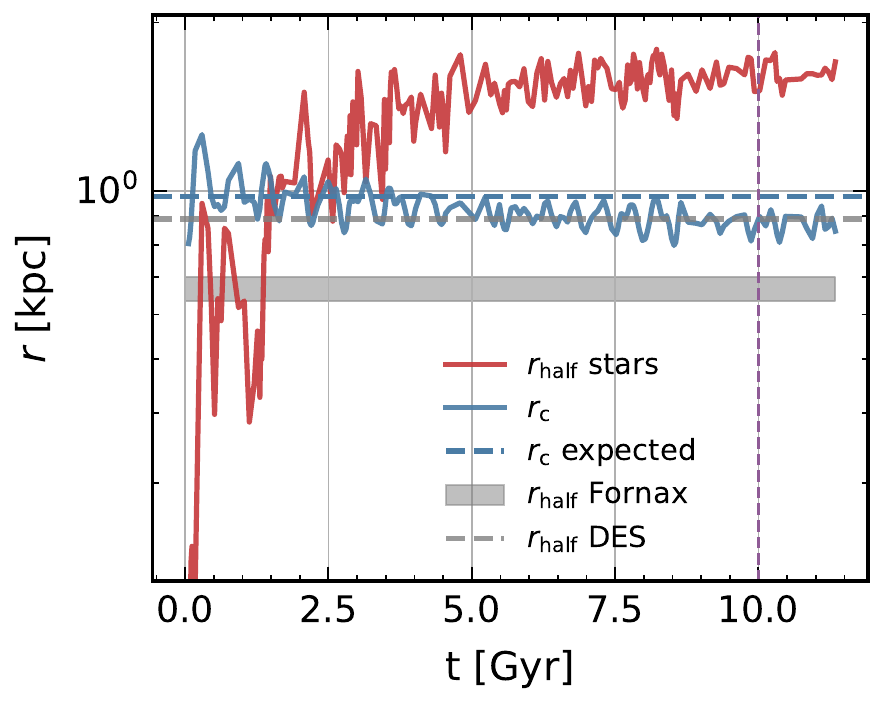}
    \caption{Simulation for a Fornax-like system for $m=\SI{1e-22}{\electronvolt}$, with $L=\SI{40}{\kilo\parsec}$. Panels explanation is the same as in the caption of Fig.~\ref{fig:NFW_1em21_main}. The rapid increase of stellar $r_{\rm half}$ makes matching both stellar kinematics and surface brightness profile difficult for a $\SI{10}{\giga\years}$ old system. 
    }
    \label{fig:1em22_large_main}
\end{figure*}

This paper is structured as follows. We first summarize the simulations and describe the relevant output quantities in Sec.~\ref{s:sim}. Next, we compare the simulation results with observational data in Sec.~\ref{s:res}. We then discuss the constraints derived from our simulations and compare them with previous works in Sec.~\ref{s:disc}. We conclude with remarks on caveats and future directions in Sec.~\ref{s:sum}. 

\section{Simulations} \label{s:sim}
We simulate halos resembling the Fornax, Leo II, and Carina dwarf galaxies. For Fornax we use $M_{200} \approx 10^9 M_{\odot}$, in a box of length $L=\SI{40}{\kilo\parsec}$ ($L=\SI{12}{\kilo\parsec}$) for $m = \SI{1e-22}{\electronvolt}$ ($m \gtrsim \SI{1e-21}{\electronvolt}$) respectively. For Carina/Leo II, $M_{200} \approx 10^8 M_{\odot}$, in a box of length $L=\SI{5}{\kilo\parsec}$ $(L= \SI{4}{\kilo\parsec})$ for $m = \SI{5e-21}{\electronvolt}$ $(m=\SI{1e-20}{\electronvolt})$. We initialize the ULDM field to a halo with either a cusp NFW~\cite{Navarro:1996gj} or a cored Burkert~\cite{Salucci:2000ps,Burkert:2015vla} profile, and the stars to a Plummer profile~\cite{Plummer1911, Dejonghe1987}, using a version of the Eddington procedure (see App.~\ref{s:Jeans}).  
The ULDM field, together with the stars, are evolved numerically for about $\SI{10}{\giga\years}$, the typical age of the stellar population~\cite{Mighell1996, de_Boer_2012, Mighell1997}. 
The ULDM is evolved using a pseudo-spectral SPE solver, whereas stars are evolved via a second-order leap-frog algorithm.  

We run a suite of simulations, varying the initial conditions of the halo and the stellar distribution, in an attempt to make the stellar LOSVD and surface brightness in the simulation as close as possible to observations. 
Additional details of the simulation set-up can be found in App.~\ref{s:Jeans} and~\ref{s:code}. 

We initialize $10^4$ stars in all our simulations. The stars are treated as test particles, and their gravitational potential is neglected. This approximation is a caveat that we discuss further below, and in App.~\ref{s:selfgravity_caveat}.

\begin{figure*}[ht]
    \centering
\includegraphics[width=0.48\linewidth]{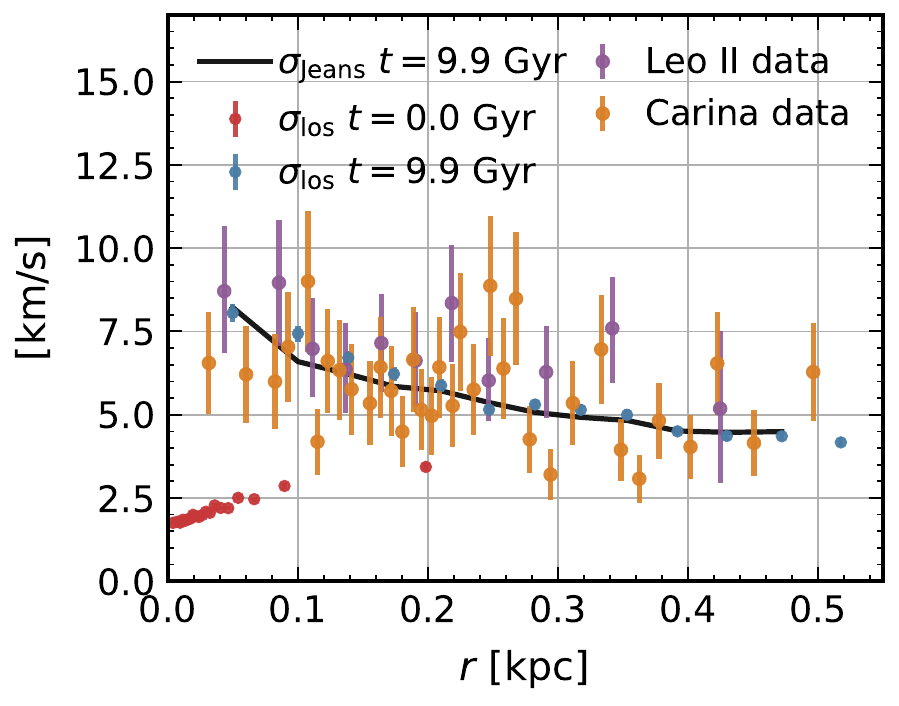}
    \includegraphics[width=0.48\linewidth]{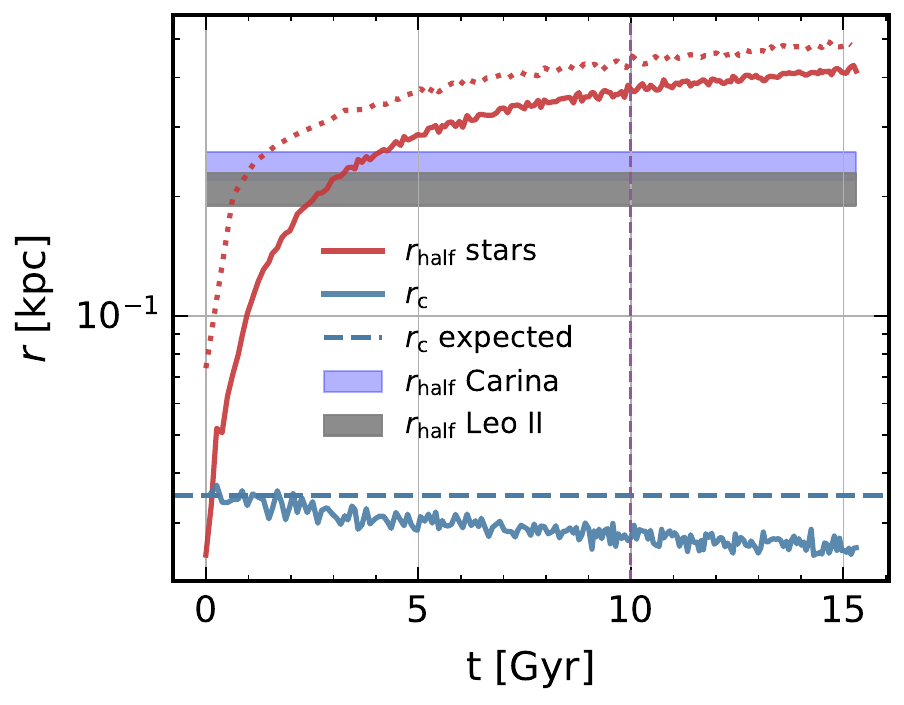}
    \caption{Simulation of a halo resembling Leo II and Carina, with $m=\SI{5e-21}{\electronvolt}$, $L=\SI{5}{\kilo\parsec}$. Legend as in Fig.~\ref{fig:NFW_1em21_main}. The stellar kinematics analysis in the left plot relates to the simulation with smaller $r_{\rm half}$. 
    Note the tension between simulated stellar $r_{\rm half}$ and data. For Leo II we derived LOSVD and $r_{\rm half}$ from Ref.~\cite{Koch:2007ye}, finding $r_{\rm half} = 0.21 \pm 0.02 \, \rm kpc$ with uncertainty determined via bootstrap resampling. For Carina we use data from Ref.~\cite{Walker:2009zp}; in this case $r_{\rm half} = 0.24 \pm 0.02 \, \rm kpc$.}
    \label{fig:NFW_5em21_LeoII_main}
\end{figure*}

\section{Results} \label{s:res} 
We conducted more than 50 simulations, varying the DM mass, density profile type (NFW or Burkert) and scale radius, the initial stellar half-light radius, and the initial stellar velocity anisotropy parameter $\beta$. The simulation suite is tabulated in App.~\ref{s:sim_details}. 
We compare the results of our simulations with observational data of Fornax~\cite{Walker:2009zp}, Leo II and Carina~\cite{Koch:2007ye}, which we describe in App.~\ref{s:data}. 

Fig.~\ref{fig:NFW_1em21_main} shows an example of an $m = \SI{1e-21}{\electronvolt}$ simulation compared with Fornax data.
The left panel shows the stellar LOSVD in the simulation at time zero (red points) and after $ t\approx \SI{9.9}{\giga\years} $ (blue points), compared with data (purple). The black line shows the prediction of a Jeans analysis using as input the true ULDM radially-averaged halo mass profile $M(r)$, stellar density $\rho_*(r)$, column density
$\Sigma_*(r_{\rm proj})$, and anisotropy parameter $\beta(r)$ from the simulation. 
The right panel shows the stellar half-light radius; as seen in the plot, $r_{\rm half}$ exhibits secular growth. We also show the soliton core size, which slowly evolves to become more compact, reflecting continued mass accretion~\cite{Dmitriev:2023ipv,Blum:2025aaa} according to the relation $M_{\rm sol} \propto 1/r_{\rm sol}$~\cite{Hui2017}.

In Fig.~\ref{fig:NFW_1em21_main}, the LOSVD peak in the soliton region, along with the flat tail outside the soliton region, are not compatible with the data. A solution that fits the tail, overshoots the data in the inner region, and vice-versa.
One might think that a negative anisotropy $\beta$ parameter could help in lowering the inner LOSVD peak; however, ULDM heating injects radial anisotropy in the star distribution. In App.~\ref{s:Jeans} we show an example of a simulation initialized with negative $\beta = -0.3$. As can be seen, during the simulation $\beta$ quickly evolves towards positive values, and its final configuration resembles that of a simulation with zero initial $\beta$.  
This suggests that positive $\beta$ is an attractor for stars in an ULDM halo (see also~\cite{DuttaChowdhury:2023qxg}), in agreement with the intuition that stars are ``kicked out" from the center of the system into eccentric orbits\footnote{It would be interesting to embed the ULDM heating effect we observe into a systematic study along the lines of, e.g., Ref.~\cite{2025arXiv250219475E}. Differently from that work, which studied radially-symmetric halo perturbations, ULDM-induced heating does not conserve angular momentum of the stellar population and we thus find that $\sigma_{\rm los}\times r_{\rm half}$ evolves in time.}. 

Although the stellar distribution is not stationary, due to the dynamical heating, a stationary Jeans analysis continues to describe the system fairly well\footnote{ But still with some discrepancy, especially on the outskirts. For the latter, it may be that outer regions
are the most susceptible to departures from equilibrium, since stars ending up there are the ones which were more strongly affected
by the time-dependent fluctuations of the ULDM halo.}. 
This can be understood by comparing the heating timescale, roughly estimated from the idealized case of homogeneous isotropic ULDM background~\cite{Bar-Or:2018pxz,Bar-Or:2020tys}\footnote{We are grateful to H. Kim and J. Eby for early assessment of this point.}
\begin{align} \label{eq:Theat}
\begin{aligned}
T_{\rm heat} =& \frac{3 \sigma^6 m^3}{16\pi^2 G^2\rho^2 \ln\Lambda} \approx \SI{0.48}{\giga\years} \qty(\frac{m}{\SI{1e-21}{\electronvolt}} )^3 \\ 
\times&\qty(\frac{\sigma}{\SI{10}{\kilo\meter\per\second}})^6 \qty(\frac{\SI{0.01}{M_\odot\per\parsec\cubed}}{\rho} )^2 \, ,
\end{aligned}
\end{align}
with the dynamical time of the system,
\begin{equation}
T_{\rm dyn} = \frac{R}{\sigma} \approx \SI{0.1}{\giga\years} \qty(\frac{R}{\SI{1}{\kilo\parsec}} ) \qty(\frac{\SI{10}{\kilo\meter\per\second}}{\sigma}) \ , 
\end{equation}
which shows $T_{\rm dyn}\lesssim T_{\rm heat}$ for $m=10^{-21}$~eV in a Fornax-like halo.  

Fig.~\ref{fig:1em22_large_main} shows a Fornax simulation with $m = \SI{1e-22}{\electronvolt}$. Here, the whole stellar body is inside the soliton region. It is then possible to find a soliton solution which fits the LOSVD and surface brightness data: this was a main result of Ref.~\cite{Schive:2014dra}. However, this solution is a brief unstable transient.
Because the heating timescale for this value of $m$ is short, the LOSVD grows with time and exceeds the observed range. 
The increase in $\sigma_{\rm los}$ is correlated with a growth in the stellar half-light radius $r_{\rm half}$, which is also difficult to reconcile with the data. Nevertheless, for masses $m \sim 10^{-22} \ \mathrm{eV}$, the heating effect in Fornax could be dramatically reduced by strong tidal interactions with the Milky Way, potentially making the evolution of $r_{\text{half}}$ compatible with the data~\cite{Yang:2025bae}.
 (See also the discussion of our neglect of stellar self-gravity, expanded in App.~\ref{s:selfgravity_caveat}: this is not a major effect for the stellar radius as observed today, but could become important if the profile was much more compact in the past.) 

Considering $m$ above $\SI{1e-21}{\electronvolt}$, the timescale for the formation of the soliton in a galaxy like Fornax begins to exceed $\SI{10}{\giga\years}$, reducing the soliton mass and its impact on stellar LOSVD. We also find that for $m\gtrsim\SI{5e-21}{\electronvolt}$ dynamical heating is negligible. (See App.~\ref{s:other_simulations}.) An adequate fit of the observational data of Fornax can be obtained in this case; this would be, in effect, a CDM-like model. 

The dynamical heating-induced growth of $r_{\rm half}$ becomes more pronounced for systems with lower velocity dispersion, to which we turn next. Considering Leo II and Carina, we show a sample result of our simulations in Fig.~\ref{fig:NFW_5em21_LeoII_main}, for $m=\SI{5e-21}{\electronvolt}$. Contrary (and complementary) to Fig.~\ref{fig:NFW_1em21_main} the LOSVD data in this case can be reproduced rather well; however, as expected, the smaller velocities in these systems make dynamical heating faster, causing rapid growth of $r_{\rm half}$ that is difficult to reconcile with the observed value. The tension becomes less severe at larger $m$: for $m\approx \SI{1e-20}{\electronvolt}$ (see App.~\ref{s:other_simulations}), $r_{\rm half}$ at 10~Gyr only slightly exceeds the data. We conclude that ULDM is in tension with data at least up to $m \le \SI{5e-21}{\electronvolt}$, and dedicated study may potentially extend the bound to slightly larger $m$. 

\section{Comparison with previous works} \label{s:disc}
We find it useful to compare our analysis with related works. Our bound $m \ge \SI{5e-21}{\electronvolt}$ is somewhat stronger than that found in~\cite{Bar:2018acw,Bar:2019bqz,Bar:2021kti} from disk galaxies. While we did not explore $m<1\times10^{-22}$~eV, complimentary bounds from cosmology disfavor that regime~\cite{Irsic:2017yje,Armengaud:2017nkf,Kobayashi:2017jcf,Leong:2018opi,Lague:2021frh}.

Ref.~\cite{DuttaChowdhury:2023qxg} conducted similar simulations of test-particle stars in an ULDM dwarf galaxy halo, confirming that dynamical heating leads to an increase of the half-light radius and the $\beta$ parameter of the stellar system. However, Ref.~\cite{DuttaChowdhury:2023qxg} considered only a single mass value, $m=\SI{8e-23}{\electronvolt}$ (again in significant tension with cosmological~\cite{Irsic:2017yje,Armengaud:2017nkf,Kobayashi:2017jcf,Leong:2018opi,Lague:2021frh} and other~\cite{Bar:2018acw,Bar:2019bqz,Bar:2021kti} bounds), and did not compare their results with observational data. 
In contrast, here we explored a range of $m$ (including values that are much more challenging to constrain with cosmology or other astrophysical systems), and compared our results to data, deriving an important constraint on $m$.

Ref.~\cite{Zimmermann:2024xvd} used Leo II data to perform a Jeans analysis, varying the ULDM density profile and carefully comparing the results with data from~\cite{Spencer_2017}. They infer $m \gtrsim 2.2 \times 10^{-21} \, \mathrm{eV}$, compatible with our result. However, that work was not based on live simulations but rather on an eigenstate expansion of the ULDM wave function. Consequently, Ref.~\cite{Zimmermann:2024xvd} did not provide information about secular evolution, which our study finds to be a significant effect.

Ref.\cite{Dalal:2022rmp} (see also~\cite{Marsh:2018zyw}, and related discussion in~\cite{Chiang:2021uvt}) argued for the exclusion of $m \lesssim 3 \times 10^{-19}~\rm eV$, based on dynamical heating estimates and focusing on the stellar radius and LOSVD in the dwarf galaxies Segue 1 and Segue 2, but without live simulations of the ULDM+stellar system\footnote{The halo in that work was constructed using approximate energy eigenmodes, computed from a mean-field model of the halo and evolved ignoring back-reaction between modes, which effectively shuts-off dynamical relaxation. This is in contrast to our live simulations that include all back-reaction effects.}. Our work here, based on simulations and applied to a different set of more massive classical dwarf galaxies, is consistent with the conclusions of Ref.~\cite{Dalal:2022rmp} in the range of $m$ we simulated. We note that the tiny Segue 1 and 2 could exhibit different tidal histories~\cite{Kirby:2013isa} and other systematic uncertainties, such as the presence of an intermediate-mass black hole~\cite{lujan2025darkmatterdominatedgalaxysegue}. As an aside, the caveat in our analysis, due to omitting stellar self-gravity, applies also to Ref.\cite{Dalal:2022rmp},~\cite{Marsh:2018zyw}, and~\cite{Chiang:2021uvt}.

Finally, Ref.~\cite{Chan:2025hhg}, Ref.~\cite{Yang:2025bae} raised the possibility that the ULDM halo of dwarf satellite galaxies is tidally stripped by the gravitational field of the Milky Way, possibly leaving behind just the soliton core. This could, in principle, ameliorate dynamical heating constraints for $m \simeq 10^{-22}$~eV. Either way, this scenario should not affect $m\gtrsim\SI{5e-22}{\electronvolt}$, the parametric region where our constraints are most important, because for such $m$ the LOSVD profiles of dwarf satellites require a halo and are not compatible with solitons alone, as we show in App.~\ref{s:tidal}. Furthermore, in order to entirely strip the halos, tidal interactions would need to be very strong across all the three systems we considered. This further highlights the importance of considering multiple systems (as we have begun to do in this work) in order to make ULDM constraints more robust.

 
\section{Conclusions and future directions} \label{s:sum}
DM-dominated dwarf galaxies can be used to study possible deviations from the CDM paradigm. In this work, we focused on how stellar kinematics and surface brightness data of dwarf spheroidal galaxies, in particular Fornax, Carina and Leo II, compare with an ULDM universe. 

For \(m = \SI{1e-21}{\electronvolt} \), stellar velocity dispersion data from Fornax is in tension with a soliton core, that our simulations show to be a ubiquitous prediction of the model. At \( m = \SI{1e-22}{\electronvolt} \), a stationary Jeans analysis can fit the data, but dynamical heating renders this fit transient. The resulting stellar velocity dispersion and half-light radius evolve significantly, requiring fine-tuned initial conditions to match observations. This result requires a significant change of perspective on early analyses of ULDM galaxies that assumed a stationary state.  
Although tidal stripping can ameliorate the situation for $m \simeq 10^{-22}$~eV, the tension remains for $m \gtrsim \SI{5e-22}{\electronvolt}$.

Galaxies with lower velocity dispersion and smaller scale radius are sensitive to higher values of $m$. We showed that for $m=\SI{5e-21}{\electronvolt}$, the observed half-light radius of the Carina and Leo II dwarf galaxies is difficult to reconcile with the rapid ULDM heating-driven evolution, that tends to predict larger $r_{\rm half}$. Our analysis suggests that Leo II and Carina (and other low-dispersion dwarfs) may allow to probe ULDM to even higher values of $m$. We postpone this study to future work.

It would also be of interest to study dynamical heating for different stellar population. In fact, dwarf galaxies such as Fornax host multiple chemo-dynamical components, with different ages~\cite{de_Boer_2012}. Our analysis does not attempt a chemodynamical decomposition and the inferred tension arises from the overall size/dispersion structure rather than from the detailed partition into sub-populations. Nevertheless, explicitly modeling the metal-rich and metal-poor components simultaneously could provide a more stringent, differential test (e.g.\ different spatial concentrations and formation times correspond to different effective exposure to ULDM-induced fluctuations).

A caveat in our analysis is that in our simulations, stars are treated as test particles. This leaves open the possibility, especially relevant if the stellar body was significantly more compact in the past, that stellar self-gravity and stellar feedback on the ULDM halo could mitigate the increase in $r_{\rm half}$. Further investigation of this issue is left for future work. We also emphasize that our constraints are based on present-day stellar structure and kinematics modeled as a single tracer population, and our analysis effectively probes the late-time configuration of the inner halo. This should be a pretty good approximation for the masses of interest, because the ULDM soliton forms/relaxes on timescales much shorter than relevant cosmological ones. Therefore, hierarchical growth or baryonic feedback effects, which predominantly affect the outer halo first and/or act most strongly during early gas-rich phases, should not qualitatively alter our conclusions. Nevertheless, a fully cosmological (and baryonic) treatment is desirable, as it could refine the detailed mapping between galaxy assembly history and the present-day inner potential, although coming with its own systematics.

Finally, our work only addressed ULDM composed of a single spin-0 field. It could be interesting to extend the analysis to higher spin (bosonic) fields, in which case the added number of degrees of freedom may reduce quasi-particle mass and thus reduce dynamical heating.

\vspace{0.2cm}

\textit{Acknowledgments.} We are grateful to J. Eby and H. Kim for their contributions to an early stage of this work. We thank J. Read for clarifications about the analysis in Ref.~\cite{Zimmermann:2024xvd} and D. Blas for useful comments.
We are grateful to E. Hardy and M. Gorghetto for useful discussion and guidance in the use of the pseudo-spectral solver. 
We thank the anonymous referees for their comments and remarks, which significantly improved this work.
The authors acknowledge the support by the European Research Area (ERA) via the UNDARK project (project number 101159929). AC is also supported by an ERC STG grant (``AstroDarkLS'', grant No. 101117510).

\appendix
\section{Eddington and Jeans analysis} \label{s:Jeans}
We initialize halos via the Eddington procedure~\cite{Widrow:1993qq,Lancaster:2019mde}
\begin{equation}
	\psi(x) = (\Delta v)^{3/2}\sum_{\vec{v}} \sqrt{f(x,\vec{v})} \e^{\iu m_i\vec{x}\cdot\vec{v} + \iu \varphi_{\vec{v}}} \ ,
\end{equation}
where $ \varphi_{\vec{v}} $ is a random phase dependent on $ \vec{v} $, $ \Delta v $ is the velocity spacing allowed by resolution in the simulation, and~\cite{2008gady.book.....B}
\begin{equation}
f(x,\vec{v}) = f_1(\mathcal{E}(x,v)) L^{-2\beta_0} \ , \ L = xv\sin\eta \ ,
\end{equation}
where $\beta_0$ is a constant anisotropy parameter, $\eta$ is the angle between the $\vec{x}$ and $\vec{v}$ vector, and 
\begin{equation}
\mathcal{E} = \Psi(r) - \frac{v^2}{2}  \ , \ \Psi = -\Phi + \Phi(r_{\rm max})  \ ,
\end{equation}
where $ \Phi $ is the gravitational potential. The value of $r_{\rm max}$ is chosen to be a few times the length of the simulation box.

If $-0.5<\beta_0<0.5$, one can write~\cite{Lacroix:2018qqh}
\begin{align} \label{eq:f_1}
\begin{aligned}
	f_1(\mathcal{E}) &= \frac{2^{\beta_0-\frac{3}2}\Gamma(\frac{3}{2}-\beta_0) \sin((\frac{1}{2}-\beta_0)\pi) }{\pi^{\frac{5}{2}} (\frac{1}{2} -\beta_0) \Gamma(1-\beta_0) } \\ 
 \times&\dv{\mathcal{E}} \int_0^{\sqrt{\mathcal{E}}} \dd{\Psi} \dv{(r^{2\beta_0}\rho)}{\Psi}{}(\mathcal{E}-\Psi)^{-\frac{1}{2} +\beta_0}\ ; 
 \end{aligned}
\end{align}
In the case of $\beta_0=0$, Eq.~\eqref{eq:f_1} reduces to
\begin{align}
	f_1(\mathcal{E}) = \frac{2}{\sqrt{8}\pi^2} \qty(\frac{1}{2\sqrt{\mathcal{E}}} \dv{\rho(\Psi = 0)}{\Psi} +  \int_0^{\sqrt{\mathcal{E}}} \dd{Q} \dv{^2\rho}{\Psi^2}{(Q)} )\ ,
\end{align}
where $  Q = \sqrt{\mathcal{E}-\Psi} $.

\begin{figure}
    \centering
\includegraphics[width=0.99\linewidth]{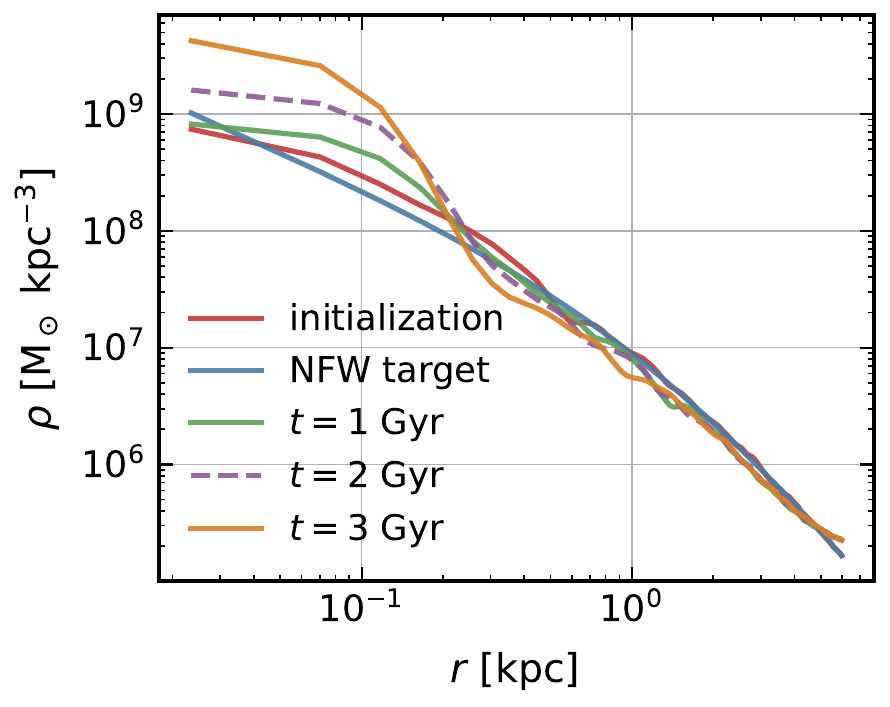}
    \caption{ Example of radially averaged density profile (with $r$ measured from the maximum density point) of a NFW profile, designed to roughly match Fornax at $m=1\times10^{-21}$~eV. We show different snapshots, including initialization (thick red) together with the target NFW density profile (thick blue) and the moment at which star particles are inserted, $t=2$~Gyr (dashed purple). The dynamical formation and growth of the soliton is manifest.}
    \label{fig:init_density}
\end{figure}

In standard N-body simulations, the Eddington procedure produces a halo that is statistically stationary up to two-body relaxation effects, which lead to slow dynamical evolution of the halo. In our ULDM simulations, the analogous effect leading to slow evolution is dynamical relaxation that can be effectively seen as two-body relaxation due to quasi-particle scattering~\cite{Bar-Or:2018pxz}. Apart from slow relaxation, however, our Eddington procedure also misses the effect of wave-mechanics pressure~\cite{Hui2017}. As a result, halos initialized via our method first react with rapid wave-pressure rearrangement on small scales on a dynamical time scale, which ``irons out" wrinkles in the density profile. This transient phase is an artifact of the scheme, not a physically motivated process. We thus ``wait it out" before populating the simulated halo with star particles. A wait time of order 1~Gyr is sufficient to remove sensitivity to initial spurious rearrangement of the halo. To be concrete, we insert star particles at $t=2$~Gyr. 
Fig.~\ref{fig:init_density} shows an example of a radially averaged density profile at different times, in particular as given  by the Eddington procedure ($t=0$) and at the time stars are inserted ($t=2$~Gyr). Notice how the characteristic soliton core can be seen in the density profile at $t\ge 1$~Gyr, while it is absent in the initial condition.

Stars are initialized using the same Eddington formalism, where $\rho$ is the density profile of stars (e.g. Plummer) and $\Phi$ is sourced by the dark matter halo. Once $f^{\rm stars}$ is computed, we can generate phase space coordinates for stars as follows. Choose a maximum radius $R_{\rm max}$ (typically, half the half length of the simulation box); then, the probability that a stars sits at a distance $x$ is
\begin{equation}
p(x) = \frac{4\pi \rho(x) x^2}{M(R_{\rm max})} \ ,
\end{equation}
and $x$ coordinates are sampled using the cumulative of $p(x)$. The probability that a star has a velocity $v$, given its position $x$, reads
\begin{equation}
p(v|x) = \frac{p(v,x)}{p(x)} = \frac{f_1(\mathcal{E}(x,v)) v^{2-2\beta_0} }{ \int\dd{v} f_1(\mathcal{E}(x,v)) v^{2-2\beta_0}  } \ ,
\end{equation}
and again $v$ coordinates are sampled via its cumulative. 
For the angle $\eta$, we use
\begin{equation}
p(\eta) = \frac{\sin^{1-2\beta_0}\eta}{\int\dd\eta\sin^{1-2\beta_0}\eta } \ .    
\end{equation}
The $\theta, \phi$ angle for the $\vec{x}$ spherical coordinates are sampled uniformly on the unit sphere; the components of the vector $v$ in spherical coordinates then read
\begin{equation}
v_r = v\cos\eta \ , \ v_\theta = v_r\cos\psi \ , \ v_\phi = v_r\sin\psi \ ,
\end{equation}
where $\psi $ is sampled uniformly on the unit circle. Then the Cartesian coordinates of $\vec{v}$ are found via
\begin{equation}
\begin{pmatrix}
v_x \\
v_y\\
v_z
\end{pmatrix} =
\begin{pmatrix}
\sin\theta \cos\phi & \cos\theta\cos\phi & -\sin\phi \\
\sin\theta \sin\phi & \cos\theta\sin\phi & \cos\phi \\
\cos\theta & -\sin\theta & 0
\end{pmatrix} 
\begin{pmatrix}
v_r \\
v_\theta\\
v_\phi
\end{pmatrix} \ .
\end{equation}

\begin{figure}[H]
\centering
\includegraphics[width=0.85\columnwidth]{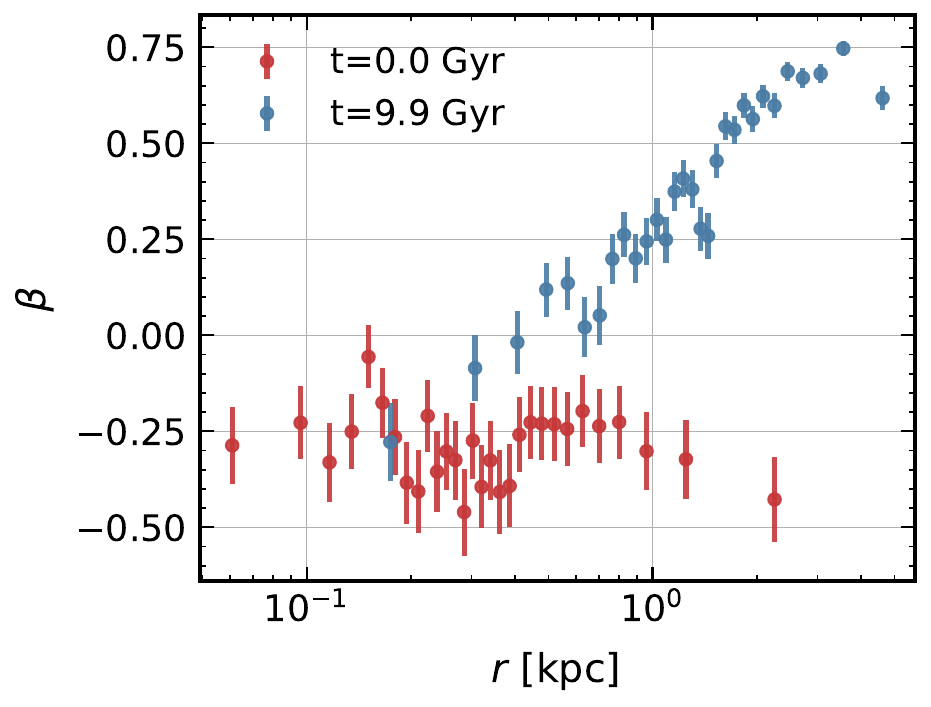}
\caption{Evolution of the anisotropy parameter for a simulation of a halo initialized as NFW, with $L=\SI{12}{\kilo\parsec}$, $m=\SI{1e-21}{\electronvolt}$, and initial $\beta = -0.3$. As noted in the main text, $\beta$ evolves towards positive values due to ULDM dynamical heating.}\label{fig:NFW_1em21Beta}
\end{figure}

\begin{figure}[H]
\centering
\includegraphics[width=0.85\columnwidth]{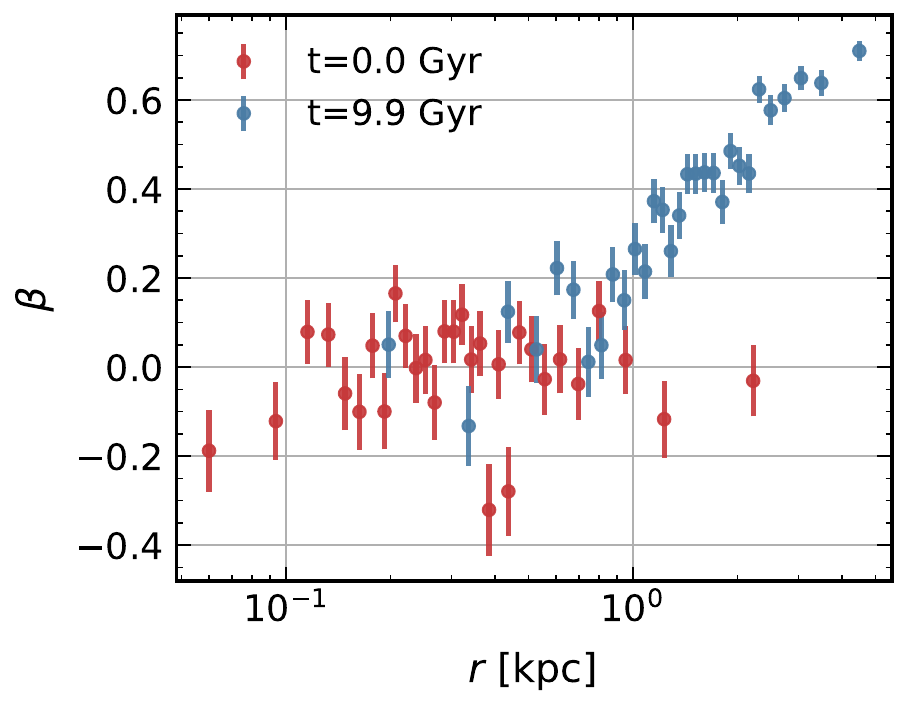}
\caption{Same as in Fig.\ref{fig:NFW_1em21Beta} but with initial $\beta = 0$.
}\label{fig:NFW_1em21Beta0ini}
\end{figure}

We have the simulation data $\vec{r}_i(t), \vec{v}_i(t)$ for the stars (labeled with the subscript $_i$), where $\vec{r}$ is computed relative to the center of mass of the stars. Given a time $t$ at which we want to compute velocity dispersions, we consider only the stars for which $\max\limits_{t'<t}(|\vec{r}_i(t')|) < R_{\rm max}$, to avoid boundary artifacts.
We then choose a random line of sight $\hat{l}$, we look for stars at a certain distance $r_{\rm proj}$, where 
\begin{equation}
r_{i,\rm proj} = |\vec{r}_i  - (\vec{r}_i \cdot \hat{l}) \hat{l}| \ ,
\end{equation} 
is the distance projected on the surface perpendicular to the random line of sight, and compute the variance of the projected velocity $\vec{v}_i\cdot\hat{l}$ of all the stars falling into the shell of thickness $\dd{r_{\rm proj}} $. This translates into
\begin{equation} \label{eq:sigma_exp}
\sigma_{\rm los} (r) = \sqrt{\mathrm{Var}[\vec{v}(r)\cdot\hat{l}]} \ .
\end{equation}

When comparing $\sigma_{\rm los} (r)$ with the result predicted by a spherically symmetric Jeans analysis, we numerically compute $\rho_*(r)$, the radially averaged star density, $\Sigma_*(r_{\rm proj})$, the star column density along the line of sight $\hat{l}$ radially averaged over the surface perpendicular to $\hat{l}$, and $M(r)$, the dark matter mass inside the radius $r$. We then use the radial velocity dispersion
\begin{equation} \label{eq:rhosigma}
\rho_*(r)\sigma^2_r(r) = G \int_r^{R_{\rm max}} \dd{s}  \frac{\rho_*(s) M(s)}{s^2} \e^{2\int_r^s \dd{z}\beta(z)/z} \ ,
\end{equation}
where $\beta(r)$ is the anisotropy parameter, defined as
\begin{equation}
    \beta(r) = 1 - \frac{\sigma^2_\theta(r) + \sigma^2_\varphi(r)}{2\sigma^2_{\rm r}(r)} \ ,
\end{equation}
where $\sigma^2_\theta(r)$, $\sigma^2_\varphi(r)$ are the tangential and azimuthal dispersion. 
Finally, the line of sight velocity dispersion reads
\begin{align} \label{eq:sigmalos}
\begin{aligned}
    &\sigma^2_{\rm los,Jeans}(r_{\rm proj}) = \frac{2}{\Sigma_*(r_{\rm proj})} \int_0^{\sqrt{R^2_{\rm max} - r^2_{\rm proj}}} \dd{Q} \\ 
& \ \times \qty(1 - \beta \frac{r_{\rm proj}^2}{Q^2 + r_{\rm proj}^2}) (\rho_*\sigma^2_r)(\sqrt{r_{\rm proj}^2 + Q^2}) \ .
\end{aligned}
\end{align}

Our simulations show that $\beta$ tends to become positive on a dynamical heating timescale. Figs.~\ref{fig:NFW_1em21Beta} and~\ref{fig:NFW_1em21Beta0ini} show a sample halo initialized as an NFW with $L=\SI{12}{\kilo\parsec}$ and $m=\SI{1e-21}{\electronvolt}$. In Fig.~\ref{fig:NFW_1em21Beta} we initialize $\beta=-0.3$, while for comparison, in Fig.~\ref{fig:NFW_1em21Beta0ini} we initialize $\beta=0$. As can be seen, both simulations evolve and reach similar values of $\beta\approx+0.5$ by $t\approx10$~Gyr.  

As remarked in the main text, the Jeans analysis using Eq.~\eqref{eq:sigmalos} reasonably matches with the $\sigma_{\rm los}$  from the simulation, Eq.~\eqref{eq:sigma_exp}. This allows, with hindsight, to use directly Eq.~\eqref{eq:sigmalos} to inform the parameters of our simulations. In particular, one can check the sensitivity to initial conditions directly by fitting the velocity dispersion data via Eq.~\eqref{eq:sigmalos}.

\begin{figure*}
    \centering    
\includegraphics[width=0.49\textwidth]{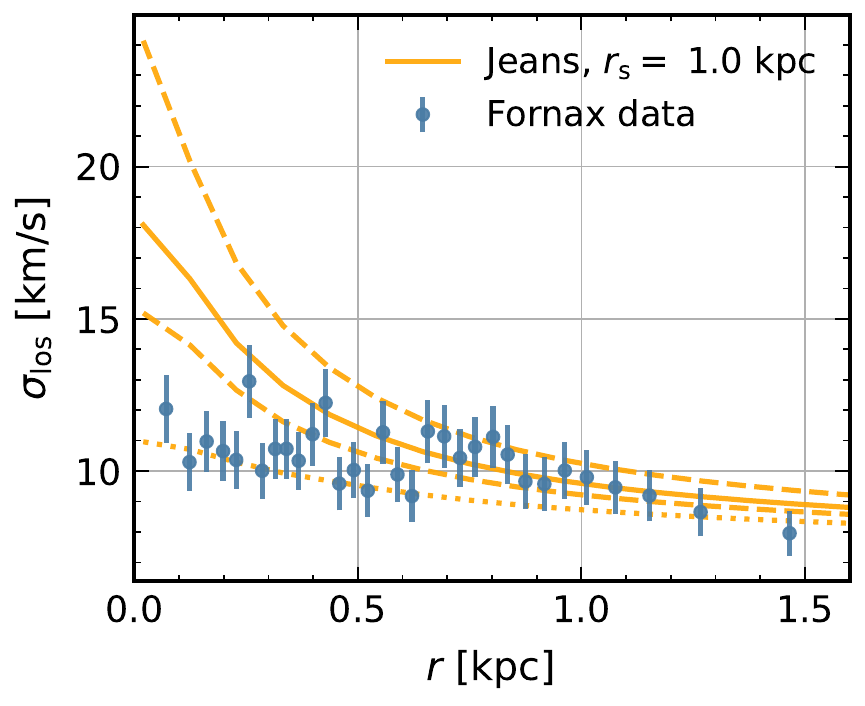}
\includegraphics[width=0.49\textwidth]{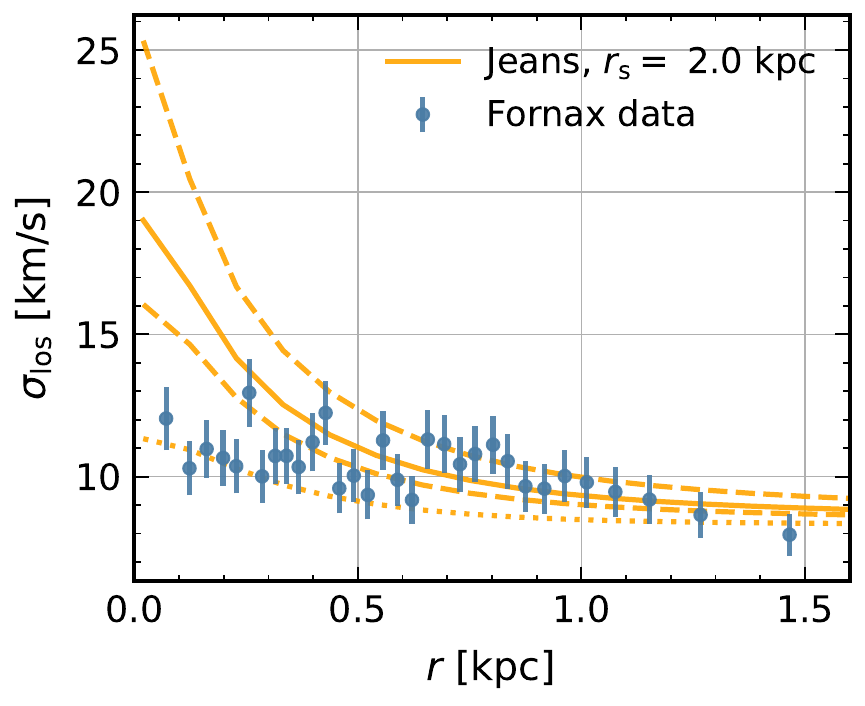}
\caption{Jeans analysis for Fornax, $m = \SI{1e-21}{\electronvolt}$. We use Eq.~\eqref{eq:sigmalos} with $\beta=0$, and a simplified mass model comprised of an NFW halo and a soliton. For the stars, we use a Plummer profile with $r_{\rm plum} = 0.85$ kpc. 
Solid line shows the result for a soliton determined via the soliton-halo relation of Ref.~\cite{Schive:2014hza}. Dashed lines relate to a soliton $1.5$ and $0.8$ more/less massive. Dotted line relates to a soliton that is $0.5$ times less massive than the soliton-halo relation prediction. A $0.5$ downward fluctuation is not seen in our 16 different Fornax $m = \SI{1e-21}{\electronvolt}$ simulations, and the results of Ref.~\cite{Yang:2025bae} suggest that it is not expected even when strong tidal stripping is implemented (moreover notice that Ref.~\cite{Yang:2025bae} discusses $m\sim \SI{e-22}{\electronvolt}$, where the soliton is more extended and hence its modifications due to tidal stripping of the outer part of the halo are expected to be more significant than for $m=\SI{e-21}{\electronvolt}$). 
The left and right panels have different values of $r_{\rm s}$ (compensated by different $\rho_{\rm s}$, as needed to roughly match the LOSVD at large radius), illustrating the in-sensitivity against variations in the NFW model parameters.}
\label{fig:Jeans}
\end{figure*}

As examples of using Eq.~\eqref{eq:sigmalos} to explore the sensitivity of stellar kinematics to halo parameters, in Fig.~\ref{fig:Jeans} we show Jeans curves for different $r_{\rm s}$, with $\rho_{\rm s}$ adjusted so that the Jeans curve fits the stellar kinematics data. The mass model has a soliton plus a NFW halo. Given the parameters $r_{\rm s}$, $\rho_{\rm s}$ of the NFW halo, we determine the mass (or equivalently, the core radius $r_{\rm c}$) of the soliton via the soliton-halo relation of Ref.~\cite{Schive:2014hza}. Dashed orange curves correspond to the same Jeans curve, with the mass of the soliton a factor 0.8 or 1.5 different with respect to the one predicted by Ref.~\cite{Schive:2014hza} relation, to explore the sensitivity of our results to different soliton-halo relations. The chosen range of soliton-halo diversity matches the one we obtain in our simulations suite for $m=10^{-21}$ eV, see also App.~\ref{s:sim_details} and Fig.~\ref{fig:sol_halo}. Dotted orange shows the same Jeans curve with soliton which is $50 \% $ less massive than expected. We remark that we do not expect to exactly obtain stellar kinematics as shown in Fig.~\ref{fig:Jeans}, since for ULDM we do not expect $\beta=0$. In particular, ULDM dynamical heating tends to increase $\beta$, which increases the central peak.   

\begin{figure}
    \centering
\includegraphics[width=0.99\columnwidth]{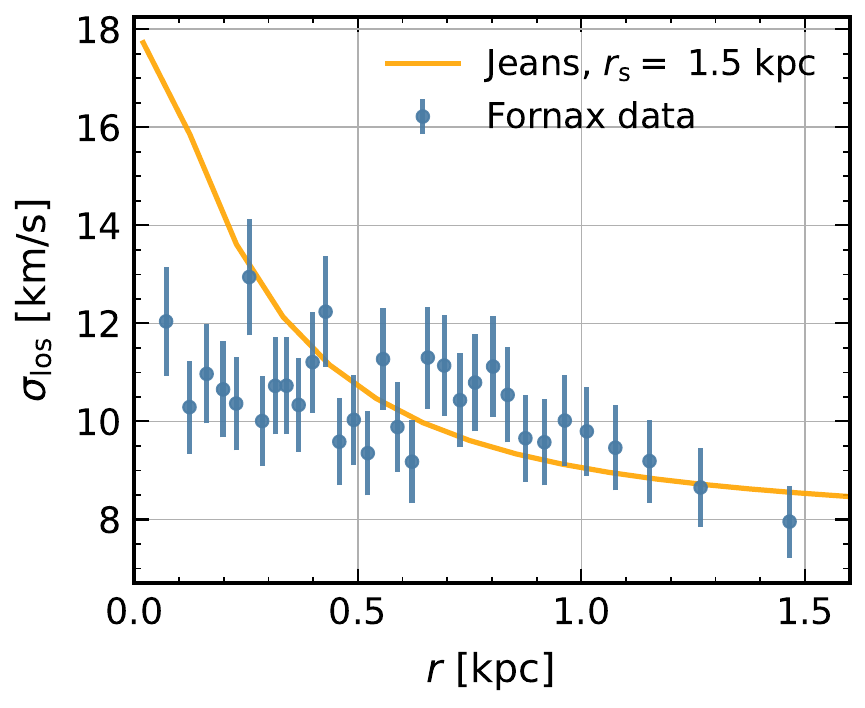}
    \caption{Jeans analysis with $r_{\rm plum} = 0.7 r^{\rm Fornax}_{\rm half}$. Other model parameters as in Fig.~\ref{fig:Jeans}.}
    \label{fig:Jeans_low_rhalf}
\end{figure}

As a of the sensitivity of the LOSVD analysis to $r_{\rm half}$, in Fig.~\ref{fig:Jeans_low_rhalf} we show a similar exercise as the one we outlined above, but for a $r_{\rm half}$ which is $0.7$ times the observed one.

\begin{figure}
    \centering
    \includegraphics[width=0.99\columnwidth]{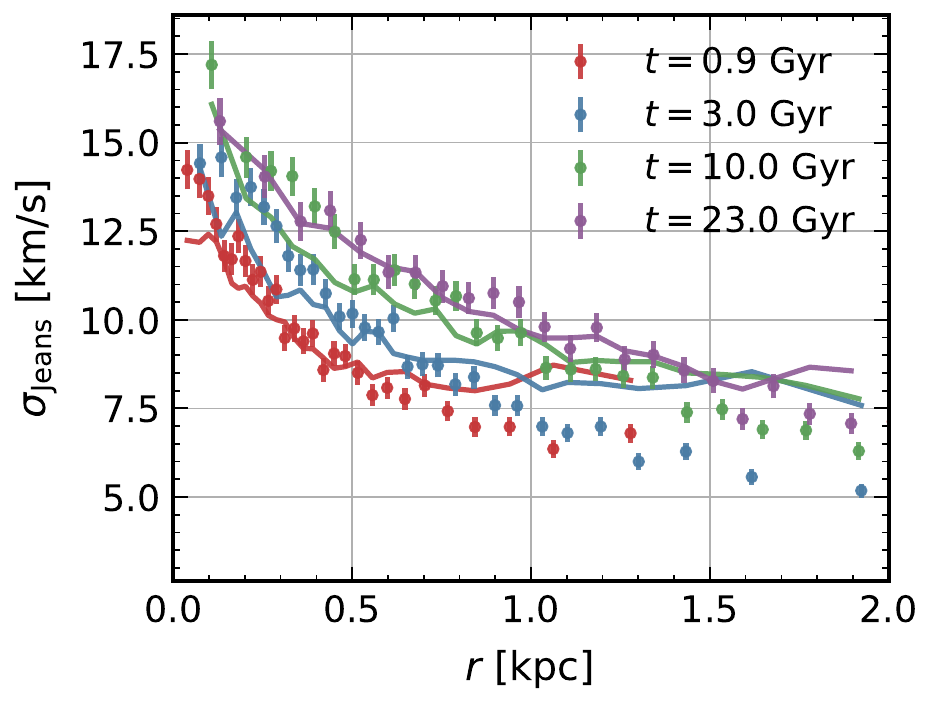}
    \caption{Stellar kinematics and Jeans curves time snapshots for the simulation showed in Fig.~\ref{fig:NFW_1em21_main}.}
    \label{fig:Jeans_time}
\end{figure}

To explore the change over time of stellar kinematics and the relative expected Jeans curves, in Fig.~\ref{fig:Jeans_time} we show stellar kinematics analysis of the system shown in Fig.~\ref{fig:NFW_1em21_main} for different time snapshots. The inner LOSVD peak forms rather early, to then slowly increase with time. The LOSVD points roughly follow the Jeans curve at all times at least for radii $r \lesssim 1$ kpc, with a better agreement for later times. 
Note that the number of stars per bin is kept constant. Consequently, at later times, the radial positions of the bins mildly shift outward due to dynamical heating, which leads to an increase in the half-light radius.

As mentioned on the main text, the agreement of the Jeans curve and simulation data in the outer regions is worse. This is probably due to greater departure from equilibrium of stars which ended up there (meaning they felt more dynamical heating than others). Moreover, the non-trivial behavior of the anisotropy parameter $\beta$ at large radii, see e.g. Fig.~\ref{fig:NFW_1em21Beta0ini}, may worsen the numerical procedure we employ to recover the Jeans curves, as interpolation of the anisotropy profile is needed to solve the integrals of Eq.~\eqref{eq:rhosigma} and Eq.~\eqref{eq:sigmalos}.

\section{ULDM Simulation code} \label{s:code}
We employ a pseudo-spectral 3D Schr\"odinger-Poisson Equation (SPE) solver, which closely resembles the one described in~\cite{Levkov:2018kau}.  Different versions of the same SPE solver implementation used in this work, which do not include star dynamics, were used in~\cite{Blum:2024igb,Gorghetto:2022sue,Budker:2023sex,Gorghetto:2024vnp}. In the following, we review the basics of its implementations.

The SPE for the field $ \psi $ with mass $ m $ read (in our conventions, $ \psi $ has the dimensions of a mass squared, in particular the density is $ \rho = |\psi|^2 $) 
\begin{align} \label{eq:SPEfield}
	&\iu\pdv{\psi}{t} = -\frac{1}{2m} \laplacian{\psi} + m\Phi\psi \ , \\
& \laplacian \Phi = 4\pi G(|\psi|^2 - \langle|\tilde\psi|^2|\rangle) \ .		
\end{align}
It is possible to re-scale fields and coordinates in dimensionless variables. To this end, define
\begin{equation} \label{eq:rescaling}
	\tilde{\psi} = \frac{1}{\lambda^2} \frac{\sqrt{4\pi G}}{m} \psi \ , \ \vec{\tilde{x}} = \lambda m \vec{x} \ , \ \tilde{t} = \lambda^2 m t \ , \
	\tilde{\Phi} = \Phi/\lambda^2 \ ,
\end{equation}
with $\lambda $ a dimensionless parameter. 
The mass $ M $, energy $ E $, and velocity $v$ transform as follows
\begin{equation}
\tilde{M} = \frac{4\pi G m}{\lambda} M \ , \ \tilde{E} = \frac{4\pi G m}{\lambda^{3}} E \ ,\ 
\tilde{v} = \frac{v}{\lambda}\ .
\end{equation}

We can then recast Eq.~\eqref{eq:SPEfield} as
\begin{align} \label{eq:SPtilde1}
	&\iu\pdv{\tilde{\psi}}{\tilde{t}} = -\frac{1}{2} \laplacian_{\tilde{x}}{\tilde{\psi}} + \tilde{\Phi} \tilde{\psi} \ , \\
	\label{eq:SPtilde2}
	& \laplacian_{\tilde{x}} \tilde{\Phi} = |\tilde\psi|^2   - \langle|\tilde\psi|^2|\rangle \ .	
\end{align}
The energy associated to the field reads 
\begin{equation}
	\tilde{E}_{\rm tot} =\tilde{E}_{\rm kin}+\tilde{E}_{\rm pot}= \int \dd[3]{\tilde{x}} \qty( \frac{|\grad{\tilde{\psi}}|^2}{2} + \frac{1}{2} |\tilde{\psi}|^2 \tilde{\Phi} ) \ .
\end{equation}

Exploiting Eq.~\eqref{eq:rescaling}, a simulation for ULDM particle mass $m$, total mass $M$, and typical velocity, length scale and time scale $v$, $x$, $t$, is equivalent to a simulation with $m'$, $M'$, $v'$, $x'$, $t'$ via the correspondence
\begin{equation} \label{eq:rescale_Leo}
 v' = \frac{1}{\lambda} v \ , \
 \frac{M'}{M} = \frac{x'}{x}= \frac{m}{\lambda m'} \ , \  t' = \lambda^2\frac{m}{m'} t \ ,
\end{equation}
with rescaling parameter $\lambda$. This feature can become useful for applying the results of one set of simulations to physical galaxy systems of different characteristic mass and size.

The ULDM field is evolved via the unitary operator
\begin{equation}
\tilde\psi(\tilde{t}+ \dd \tilde{t}) = \prod_{\alpha} \e^{-\iu d_\alpha \dd{\tilde{t}} \tilde{\Phi}_\alpha} \e^{-\iu c_\alpha\dd{\tilde{t}} \frac{(-\iu \grad)^2}{2} }\tilde\psi(\tilde{t}) \ .
\end{equation}
The equation is meant to be read from right to left, i.e. before one applies the kinetic operator
\begin{equation}\label{eq:kin_op}
\e^{-\iu c_\alpha\dd{\tilde{t}} \frac{(-\iu \grad)^2}{2} }\tilde\psi(\tilde{t}) =: \tilde\psi^{(\alpha)}(\tilde{t} +\dd{\tilde{t}} ) \ ,
\end{equation}
and then the potential operator, where
\begin{equation}\label{eq:Poiss_num}
\laplacian{\tilde\Phi_\alpha} = |\tilde\psi^{(\alpha)}|^2 \ .
\end{equation}
The constants $ c_\alpha $, $ d_\alpha $ can be found in~\cite{Levkov:2018kau}. Eqs.~\eqref{eq:kin_op},\eqref{eq:Poiss_num} are solved using fast Fourier transforms using the FFTW library~\cite{10.1145/301631.301661}. 

We use adaptive time steps, to ensure a conservation of energy $ \Delta E/E \lesssim 10^{-5} $ between time-steps, and an overall conservation of energy $ |E^{\rm tot}_{\rm final} - E^{\rm tot}_{\rm initial}|/|E^{\rm tot}_{\rm final} + E^{\rm tot}_{\rm initial}| \lesssim 10^{-3} $.

Star dynamics is computed from the gravitational potential $\Phi$, which is assumed to be sourced only from the ULDM field $\psi$, neglecting stellar gravity. Star coordinates are evolved via the leapfrog integrator~\cite{2008gady.book.....B}
\begin{align}
\begin{aligned}
    &\vec{r}_i(t_{j+1/2}) = \vec{r}_i(t_j) + \frac{1}{2} \dd{t} \vec{v}_i(t_j) \ , \\ 
    &\vec{v}_i(t_{j+1}) = \vec{v}_i(t_j) - \dd{t} \grad\Phi(\vec{r}_i(t_{j+1/2})) \ ; \\ 
    &\vec{r}_i(t_{j+1}) = \vec{r}_i(t_{j+1/2}) + \frac{1}{2} \dd{t} \vec{v}_i(t_{j+1/2}) \ .
\end{aligned}
\end{align}
$\grad\Phi(\vec{r}_i(t_{j+1/2}))$ is computed from the ULDM-sourced $\Phi$ in the grid, using multi-linear interpolation. 
Stars coordinate initialization was described in Sec.~\ref{s:Jeans}.

\begin{figure}[ht]
    \centering
\includegraphics[width=0.9\linewidth]{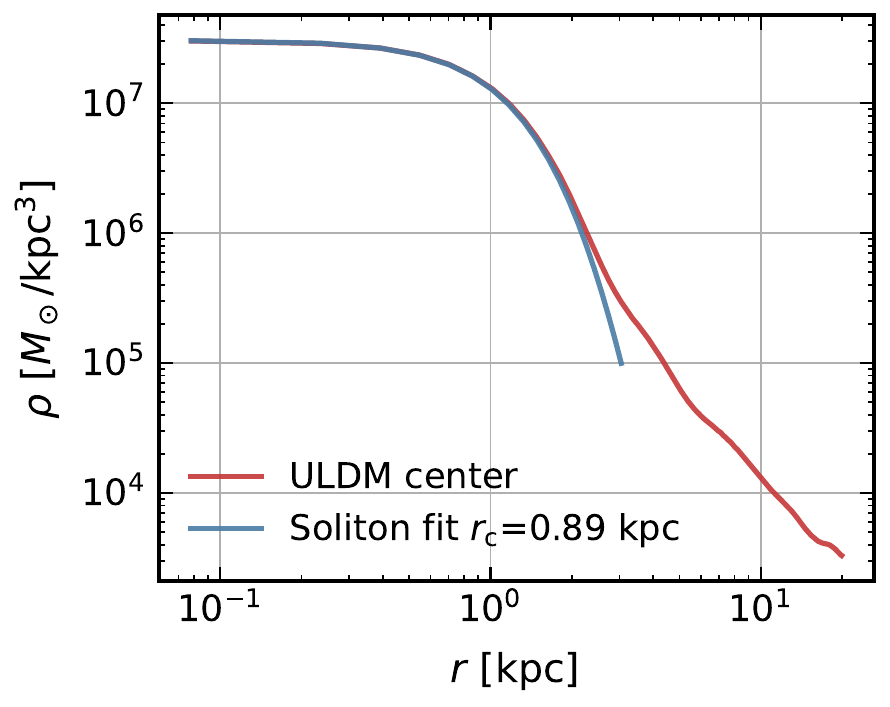}
\caption{ULDM density profile (red curve) for a simulation with $L=\SI{40}{\kilo\parsec}$, $m=\SI{1e-22}{\electronvolt}$, evolved for $10 \, \rm Gyr$. We also show (blue curve) the soliton fit using Eq.~3 of Ref.~\cite{Schive:2014hza}; the resulting core radius is $r_c = 0.89 \, \rm kpc$, close to the value found in Ref.~\cite{Schive:2014hza}, $r_{\rm expected} = \SI{0.98}{\kilo\parsec}$.}
    \label{fig:NFW_1em22_density}
\end{figure}

Our simulations generally exhibit soliton formation in good agreement with the original results of Ref.~\cite{Schive:2014hza} (at least for $m\lesssim1\times10^{-21}$~eV, when the dynamical relaxation time is not too long, as noted in the text).  Fig.~\ref{fig:NFW_1em22_density} shows an example of ULDM density profile from a simulation initialized as an NFW halo with $L=\SI{40}{\kilo\parsec}$, $m=\SI{1e-22}{\electronvolt}$, evolved for $9.9 \, \mathrm{Gyr}$. The formation of a soliton core is clearly visible, with size consistent with expectations from the soliton-halo mass relation, and a functional form (blue curve fit)
as predicted by Eq.~3 of Ref.~\cite{Schive:2014hza}.

\begin{figure*}
    \centering
    \includegraphics[width=0.48\linewidth]{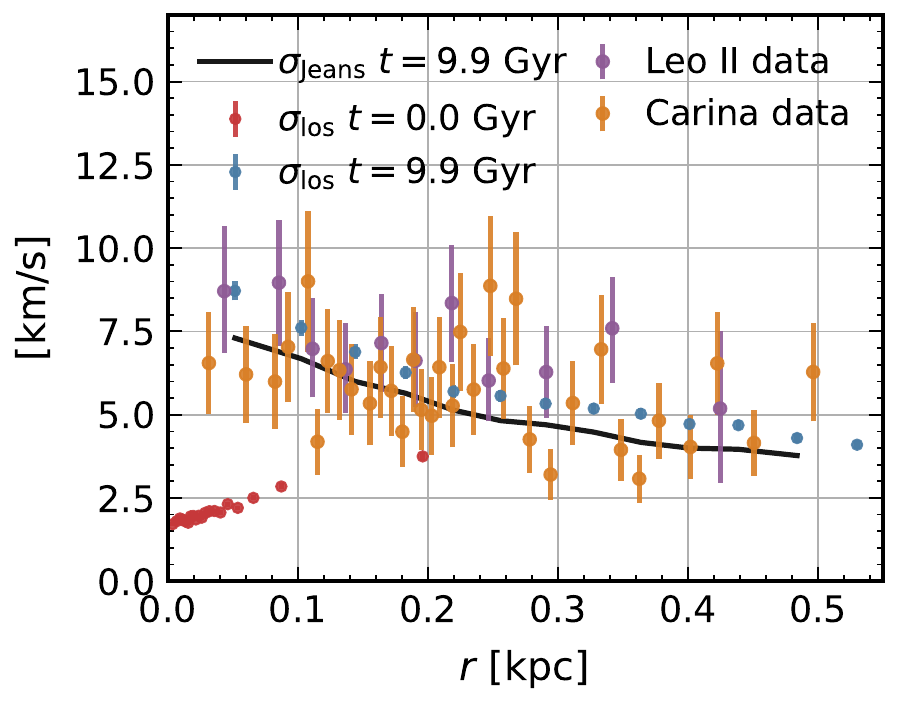}
    \includegraphics[width=0.48\linewidth]{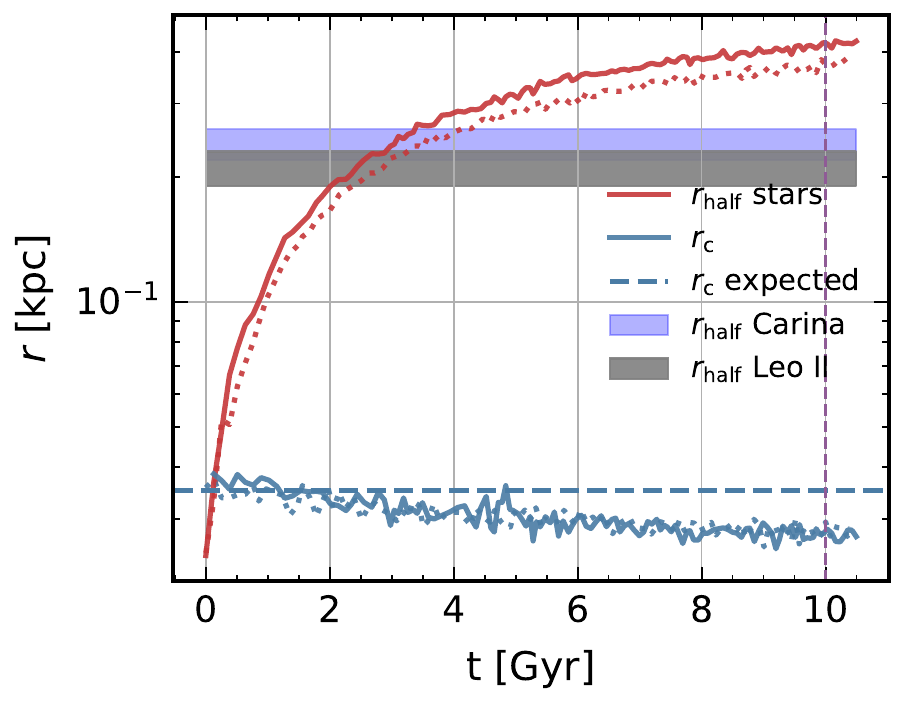}
    \caption{Same run as in Fig.~\ref{fig:NFW_5em21_LeoII_main} of the main text, but with doubled resolution ($512^3$ points compared with $256^3$ points) and different initial random seed. On right panel, dotted red line and dotted blue line refer to $r_{\rm half}$ and $r_{\rm c}$ of the run of Fig.~\ref{fig:NFW_5em21_LeoII_main}, showed here for direct comparison. Numerical results are consistent.}
    \label{fig:LeoII_highres}
\end{figure*}

As a numerical convergence test of our code implementation, in Fig.~\ref{fig:LeoII_highres} we show the same simulation as the one showed in Fig.~\ref{fig:NFW_5em21_LeoII_main}, but with different initial random seed and doubled resolution.
Results are consistent with Fig.~\ref{fig:NFW_5em21_LeoII_main}.

\section{\boldmath When is stellar self-gravity important?}
\label{s:selfgravity_caveat}

Our analysis, treating ULDM and stars dynamically in tandem and comparing the results directly to observational data, is a step forward in comparison to previous work. However, a remaining limitation is the treatment of stars as test particles. This approximation seems reasonable given the dark-to-luminous mass ratios observed in dwarf spheroidal galaxies today, but it can fail if, in the past, the stellar population was more concentrated. For Fornax, Ref.~\cite{Battaglia_2015} quotes a dark-to-luminous mass ratio of about 5--6 (16--18) within 1.6 (3) kpc~\cite{Battaglia_2015}. To illustrate how the dark-to-luminous mass ratio scales with the parameters of interest and within the radius of the galaxy, let us assume an NFW profile for DM and a Plummer profile~\cite{Plummer1911, Dejonghe1987} for the stellar distribution.

The enclosed NFW mass profile is given by
\begin{equation}
    M_{\mathrm{NFW}}(<r) 
    = 4\pi \, r_s^3 \, \rho_s 
    \Bigl[
      \ln\bigl(1 + \tfrac{r}{r_s}\bigr) 
      - \frac{r}{r + r_s}
    \Bigr],
\end{equation}
while the enclosed Plummer profile is
\begin{equation}
    M_{\mathrm{Plummer}}(<r) 
    = M^*_0 
      \Bigl(\tfrac{r}{r_{\mathrm{half}}}\Bigr)^3
      \frac{1}{\bigl(1 + (\tfrac{r}{r_{\mathrm{half}}})^2\bigr)^{3/2}},
\end{equation}
where $M^*_0$ is a normalization constant that accounts for the total number of stars, and $r_{\mathrm{half}}$ is their half-light radius.

\begin{figure}[H]
\centering
\includegraphics[width=0.9\columnwidth]{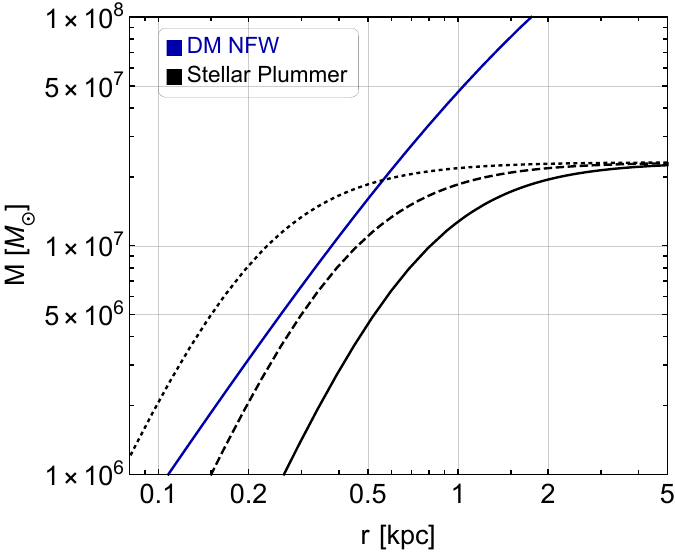}
\caption{NFW mass profile for DM with $r_s = 1.5\, \rm kpc$ and $\rho_s = 0.01 \, \rm M_\odot/pc^3$ (blue curve) and different stellar Plummer mass profiles (black curves). For the stellar profiles we have a Plummer profile with $r_{\rm half} = 0.7, 0.4, 0.2 \, \rm kpc$, for the solid, dashed, and dotted curves, respectively.}\label{fig:MassProfiles}
\end{figure}

In Fig.~\ref{fig:MassProfiles}, we show the NFW mass profile (blue solid line) for $r_s = 1.5\,\mathrm{kpc}$ and $\rho_s = 0.01\,M_{\odot}/\mathrm{pc}^3$, which fits well the LOSVD data at large radii, alongside three different stellar mass profiles (black lines). The solid black curve corresponds to a Plummer profile with $r_{\mathrm{half}} = 0.7\,\mathrm{kpc}$, chosen to match the latest determination of Ref.~\cite{2009ApJ...704.1274W} adopted in the main text, and a normalization $M^*_0$ set so that
\begin{equation}
  \frac{M_{\mathrm{NFW}}(<1.6\,\mathrm{kpc})}
       {M_{\mathrm{Plummer}}(<1.6\,\mathrm{kpc})}
  = 5.
\end{equation}
In this case DM dominates by a factor of a few inside all the radii of interest. Then, the dashed and dotted curves keep the same total stellar mass normalization (integrated to infinity) but reduce $r_{\mathrm{half}}$ to $0.4\,\mathrm{kpc}$ and $0.2\,\mathrm{kpc}$, respectively. 

We conclude that if the same number of stars were compressed by a factor $\sim 3-4$ in their radial distribution, then stellar self-gravity would become significant, and eventually dominate the dynamics, rendering our simulations inadequate in that regime. This means that our simulations of the early evolution in some of the stellar half-light radius curves in Figs.~\ref{fig:NFW_1em21_main},~\ref{fig:1em22_large_main}, and~\ref{fig:NFW_5em21_LeoII_main} are not fully consistent. The simulations should evolve into control once the stellar populations expand, within a few Gyr, and well before $t\approx10$~Gyr in all cases. 

\section{Dwarf galaxy data} \label{s:data}

For LOSVD data we have analyzed the following \href{https://cdsarc.cds.unistra.fr/ftp/J/AJ/137/3100/}{repository} from Ref.~\cite{Walker:2009zp}. This catalog has spectroscopic data for individual stars observed as part of the Michigan/MIKE Fiber System
survey of four dwarf spheroidal galaxies: Carina, Fornax, Sculptor, and Sextans. For Fornax, following~\cite{Walker:2009zp} we selected stars with membership probability of at least $95\%$. From these, we extracted stellar velocities, binned the data by radial distance from the galaxy center (ensuring each bin contained the same number of stars), and calculated the velocity dispersion for each bin.

\begin{figure}[H]
\centering
\includegraphics[width=0.95\columnwidth]{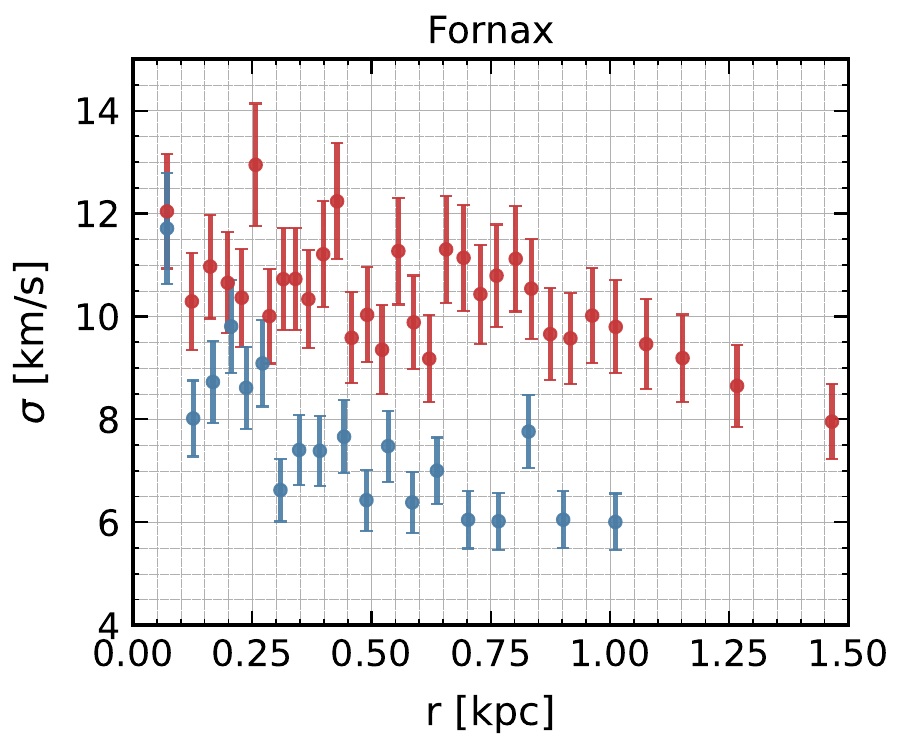}
\caption{LOSVD profile for Fornax. The red (blue) data points correspond to a $95\%$ ($99\%$) membership probability cut. 
}\label{fig:DifferentMembershipCutsFornax}
\end{figure}

We note that the choice of membership probability threshold can impact the results. Fig.~\ref{fig:DifferentMembershipCutsFornax} illustrates the LOSVD profiles for Fornax with two different membership probability thresholds: $95\%$ (red data points) and $99\%$ (blue data points). In both cases, each bin contains 60 stars. The results show that the stricter $99\%$ threshold substantially affects the inferred LOSVD profile. This cut also reduces the number of stars in the sample from 1920 to 1140.

In the main text we quote the half-light radius reported in Tab.~I of Ref.~\cite{2009ApJ...704.1274W}, $r_{\rm half} = 0.668 \pm 0.034 \,\mathrm{kpc}$, which is also based on the data from the Michigan/MIKE Fiber System survey. 
(Ref.~\cite{2009ApJ...704.1274W} has an Erratum which affects the half-light radius. See also the latest arXiv version of the paper.) 
The DES Collaboration~\cite{DES:2018jtu} found a higher value from their survey $r_{\rm half} = 0.89 \, \rm kpc$\footnote{Another study, based on older data~\cite{McConnachie_2012}, reports $r_{\rm half} = 0.710 \pm 0.077 \, \rm kpc$ (see their Table III).}, and we show that too in relevant plots for reference. It is worth noting that Table II of Ref.~\cite{DES:2018jtu} shows that $r_{\rm half}$ can vary by a factor of two between, for example, the horizontal branch and the blue plume star sub-samples.  
The data for Carina was also obtained from Ref.~\cite{Walker:2009zp} with membership probability cut of $95\%$. The half-light radius is $r_{\rm half} = 0.24 \pm 0.02 \, \rm kpc$, where the uncertainty was determined via bootstrap resampling~\cite{Efron1979}.

Finally, we have analyzed data for the Leo II dwarf spheroidal galaxy, available in this \href{https://cdsarc.cds.unistra.fr/ftp/J/AJ/134/566/}{repository}. This dataset provides the LOSVD derived from 200 stars measured using the FLAMES/GIRAFFE spectrograph at the European Southern Observatory in Chile~\cite{Koch:2007ye}. We applied the same cut as in Ref.~\cite{Koch:2007ye}, retaining only stars within $3\sigma$ of the mean radial velocity. This cut reduced the number of stars to 171. These were then binned into groups of 12 stars each, reproducing fairly well Fig.7 (left panel) from Ref.~\cite{Koch:2007ye}. Using the same data we also derived the observed half light radius for this population to be $r_{\rm half} = 0.21 \pm 0.02 \, \rm kpc$.

\section{\boldmath Fornax simulation for $m = 5 \times 10^{-21} \, \rm eV$ and $m = 10^{-22} \, \rm eV$ and LeoII/Carina for $m = 10^{-20} \, \rm eV$}
\label{s:other_simulations}

\begin{figure*}[ht]
    \centering
    \includegraphics[width=0.47\linewidth]{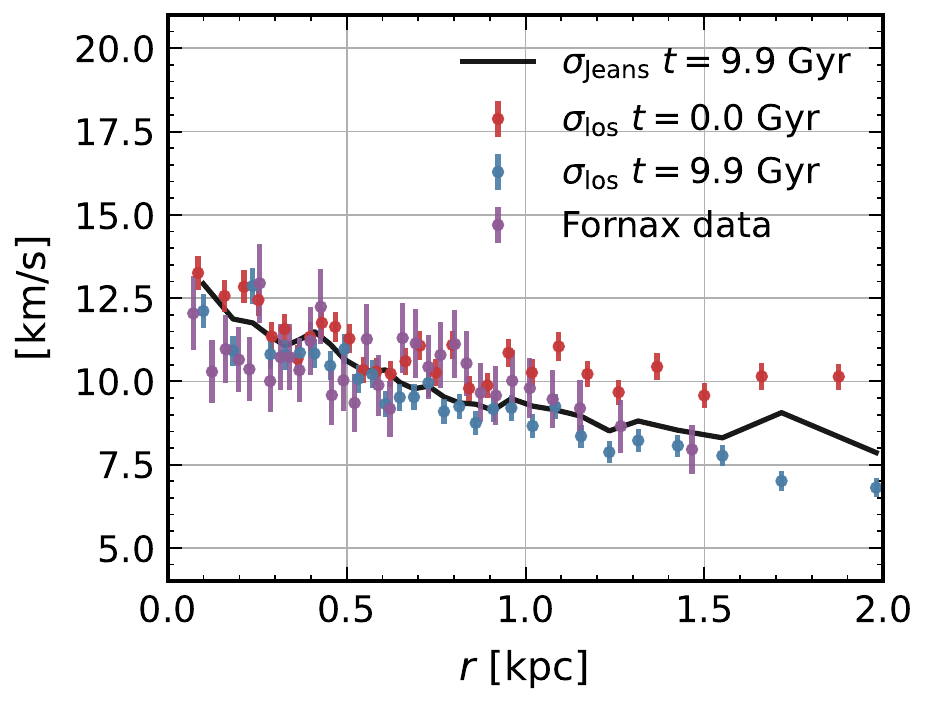}
     \includegraphics[width=0.495\linewidth]{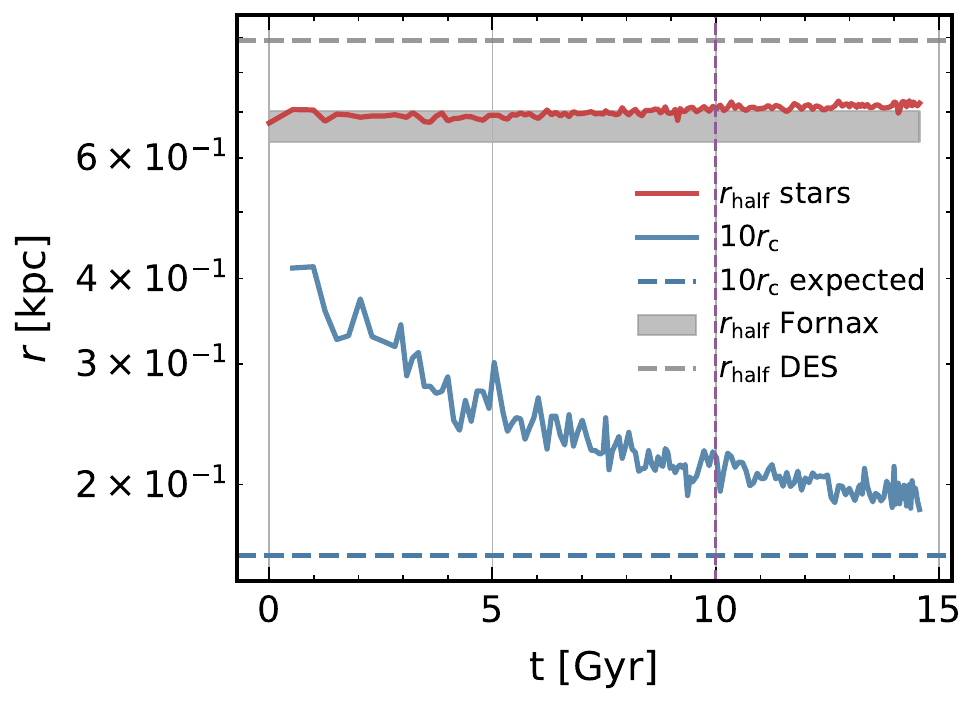}
     \includegraphics[width=0.48\linewidth]{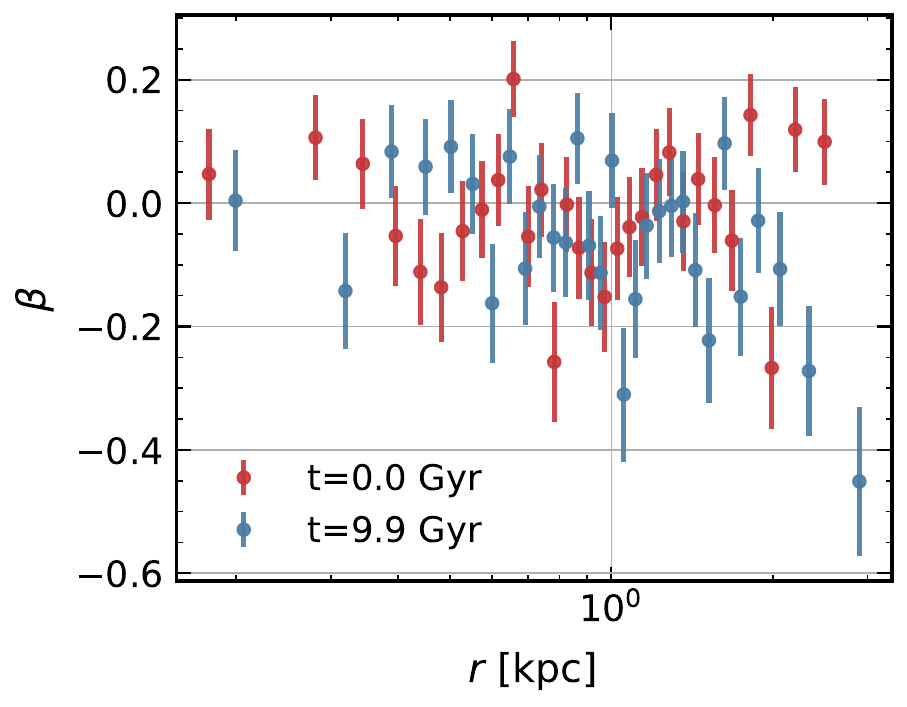}
  \includegraphics[width=0.475\linewidth]{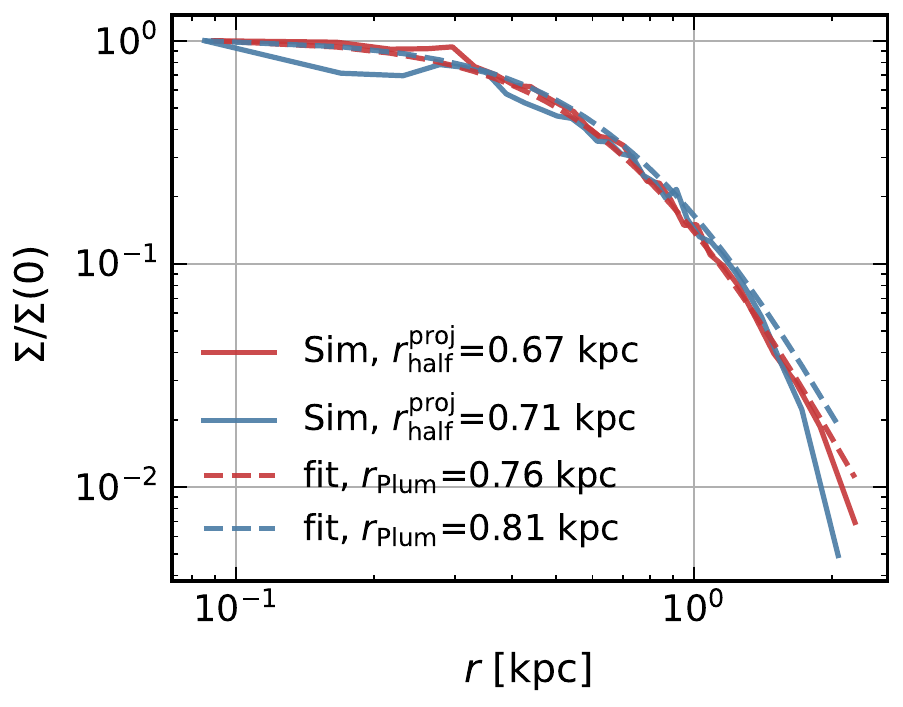}
    \includegraphics[width=0.48\linewidth]{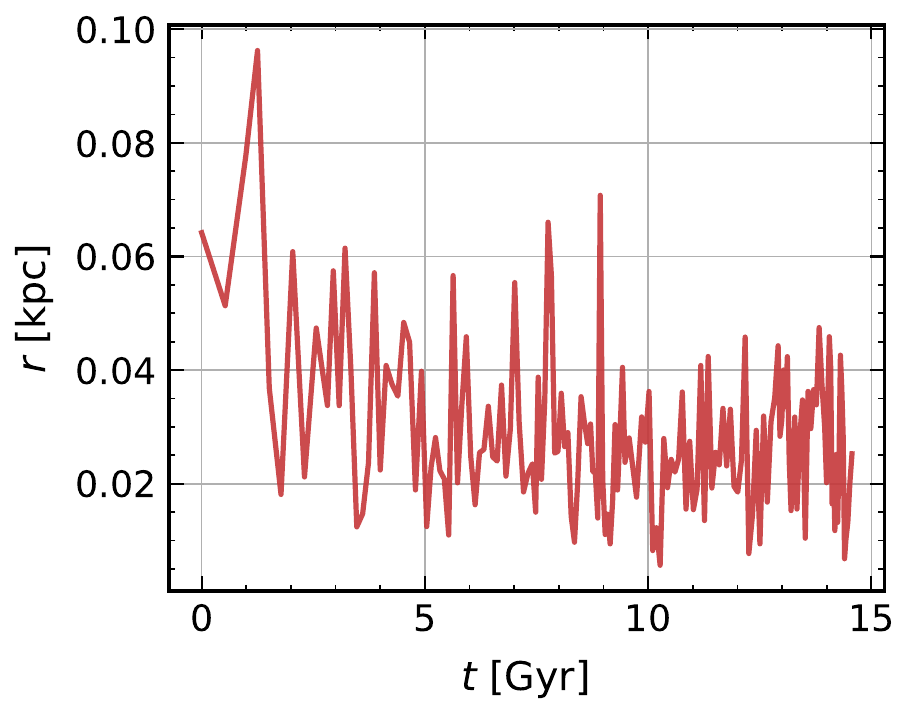}
    \caption{Simulation initialized as an NFW halo, with $L=\SI{8}{\kilo\parsec}$, $m=\SI{5e-21}{\electronvolt}$. This is an example where dynamical heating is not significant, as we approach the ``CDM limit" of ULDM.
    \textbf{Upper left.} LOSVD data for Fornax (purple) is compared to the simulation result at time 0 (red) and at time $ t\approx\SI{9.9}{\giga\years} $ (blue). Black line is the $\sigma_{\rm los}$ predicted by a Jeans analysis. 
    \textbf{Upper right.} Red line: stellar $r_{\rm half}$ evolution in time. 
    Grey band shows the observed $r_{\rm half}$ from Ref.~\cite{2009ApJ...704.1274W}, grey dashed line shows $r_{\rm half}$ from Ref.~\cite{DES:2018jtu}. Blue line shows the soliton core radius $r_{\rm c}$. 
    Dashed blue line shows the value of $r_{\rm c}$ expected from the soliton-halo relation of Ref.~\cite{Schive:2014hza}. Vertical dashed line denotes $t=10$~Gyr.
    \textbf{Center left.} Anisotropy parameter $\beta$ for $t=0$ (red) and $t=\SI{9.9}{\giga\years}$ (blue). 
    \textbf{Center right.} Radially averaged stellar column density in the simulation for $t=0$ (red) and $t=\SI{9.9}{\giga\years}$ (blue), compared with a Plummer profile fit (dashed). We quote the Plummer scale parameter of the fit $r_{\rm Plum}$, together with $r_{\rm half}^{\rm proj}$ from the simulation.
    \textbf{Bottom.} Displacement between the stars center of mass and the maximum ULDM density point, which approximately coincides with the soliton.
    }
    \label{fig:NFW_5em21}
\end{figure*}

\begin{figure*}[ht]
    \centering
      \includegraphics[width=0.48\linewidth]{plots/Fornax_1em22_Burkertjeans_sigma.pdf}
      \includegraphics[width=0.48\linewidth]{plots/Fornax_1em22_Burkertrhalf_stars.pdf}
    \includegraphics[width=0.48\linewidth]{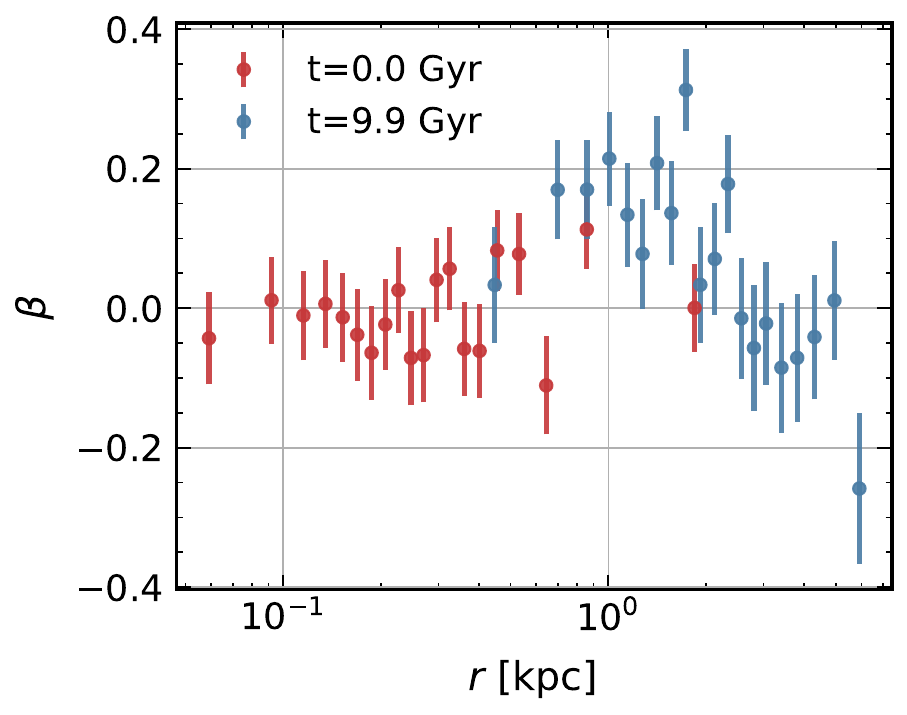}
    \includegraphics[width=0.48\linewidth]{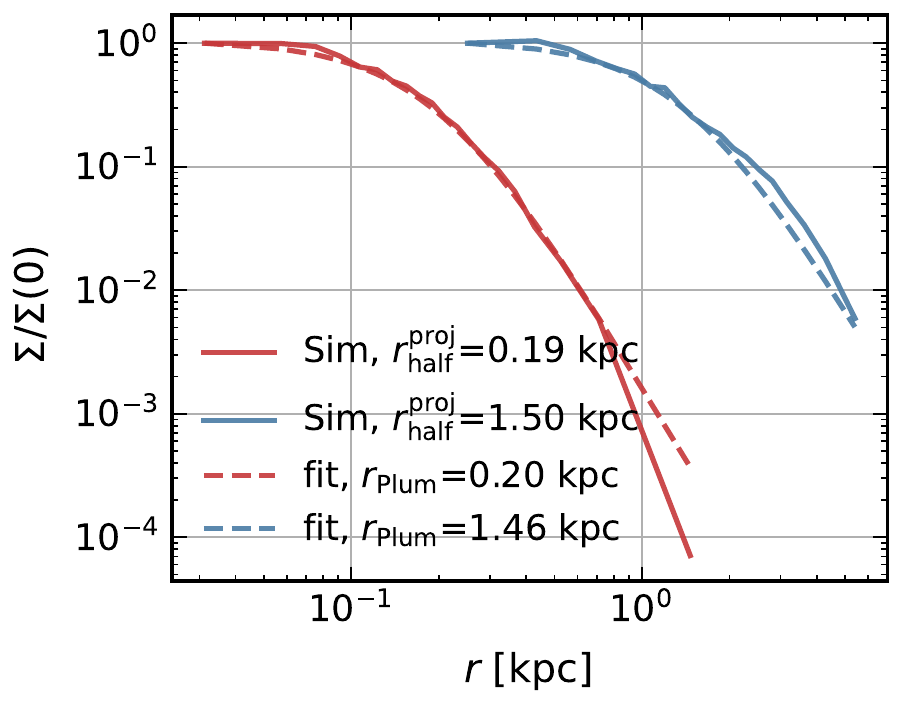}
    \includegraphics[width=0.48\linewidth]{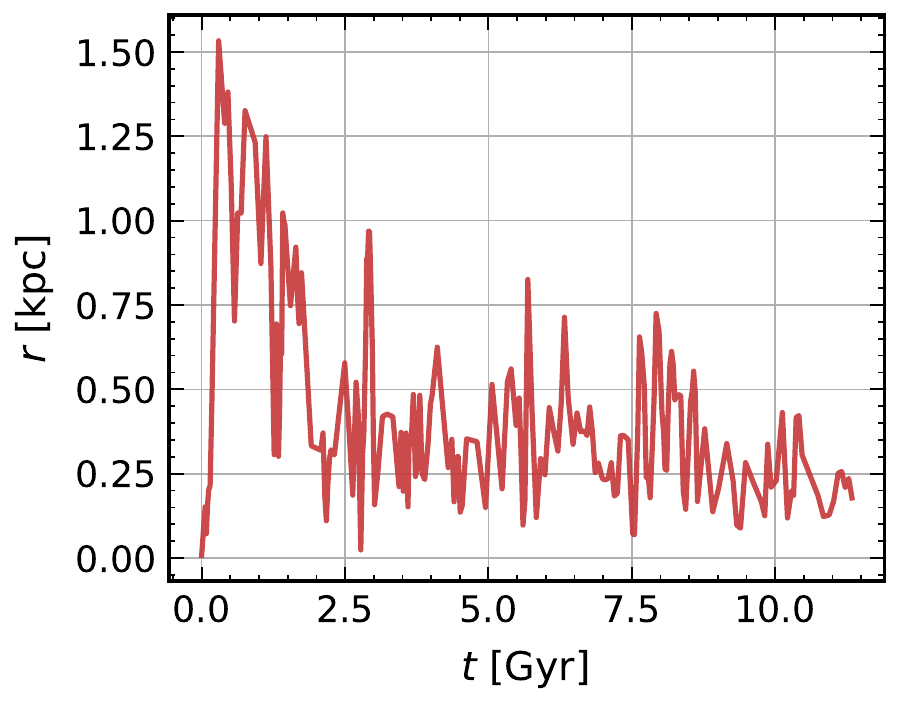}
     \includegraphics[width=0.48\linewidth]{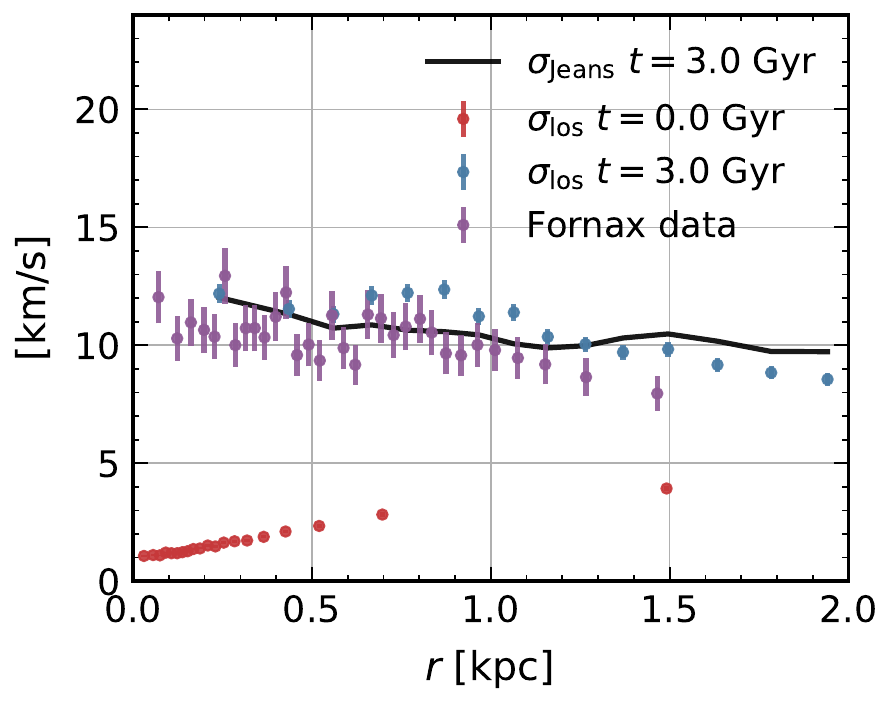}
    \caption{Simulation for a Fornax-like system for $m=\SI{1e-22}{\electronvolt}$, with $L=\SI{40}{\kilo\parsec}$. Panels explanation is the same as in Fig.~\ref{fig:NFW_5em21}, with the addition of the bottom-right panel, which is the same as the upper-left panel, but for the $t=\SI{3}{\giga\years}$ snapshot. 
    }
    \label{fig:1em22_large}
\end{figure*}

In this section we present detailed results from two simulations of Fornax: one (Fig.~\ref{fig:NFW_5em21}) conducted with $L = \SI{8}{\kilo\parsec}$, $m = \SI{5e-21}{\electronvolt}$, and an initial $\beta$ parameter set to zero in the Eddington procedure; and another  (Fig.~\ref{fig:1em22_large}) with $L = \SI{40}{\kilo\parsec}$, $m = \SI{1e-22}{\electronvolt}$, and initial $\beta \sim 0$. These  simulations represent the extremes of the $m$ range we explored.

For the case of $m = \SI{5e-21}{\electronvolt}$, ULDM heating is unimportant on the timescales of interest, and all quantities remain close to their initial values, including the stellar surface brightness profile, and the $\beta$ parameter which fluctuates around zero. 
We checked that the spread in $\beta$ around zero is consistent with random statistical  fluctuations. 

We also observe that the soliton core radius remains larger (by about a factor of 3) than would be expected from the naive soliton--halo relation. 
This suggests that dynamical relaxation is still ongoing in the system.

For the case of $m = \SI{1e-22}{\electronvolt}$, the soliton encompasses the stellar body. The heating time-scale is much shorter in this case, and $r_{\rm half}$ doubles the observed one within $ \lesssim \SI{4}{\giga\years}$. 
We emphasize that even though a soliton can naively fit Fornax data according to a Jeans analysis, dynamical heating makes $\sigma_{\rm los}$ and $r_{\rm half}$ evolve in a nontrivial way. To illustrate this point, the bottom-right panel of Fig.~\ref{fig:1em22_large}  
shows $\sigma_{\rm los}$ at an intermediate time $t=\SI{3}{\giga\years}$ through the simulation. Around that moment in time, both $\sigma_{\rm los}$ and $r_{\rm half}$ are roughly consistent with the data, compatible with the results of~\cite{Schive:2014dra}. However, again, this is a short-lived transient state, and by $t\approx10$~Gyr both $r_{\rm half}$ and $\sigma_{\rm los}$ are off.  
We note that the soliton in this case quickly relaxes close to the soliton--halo relation. 

\begin{figure*}
    \centering
    \includegraphics[width=0.48\linewidth]{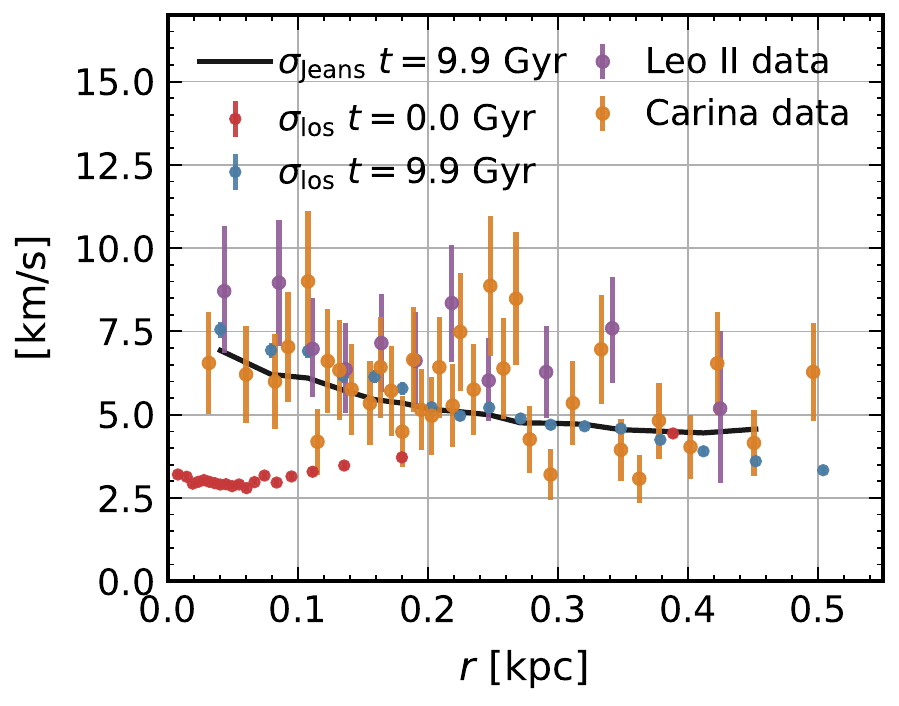}
    \includegraphics[width=0.48\linewidth]{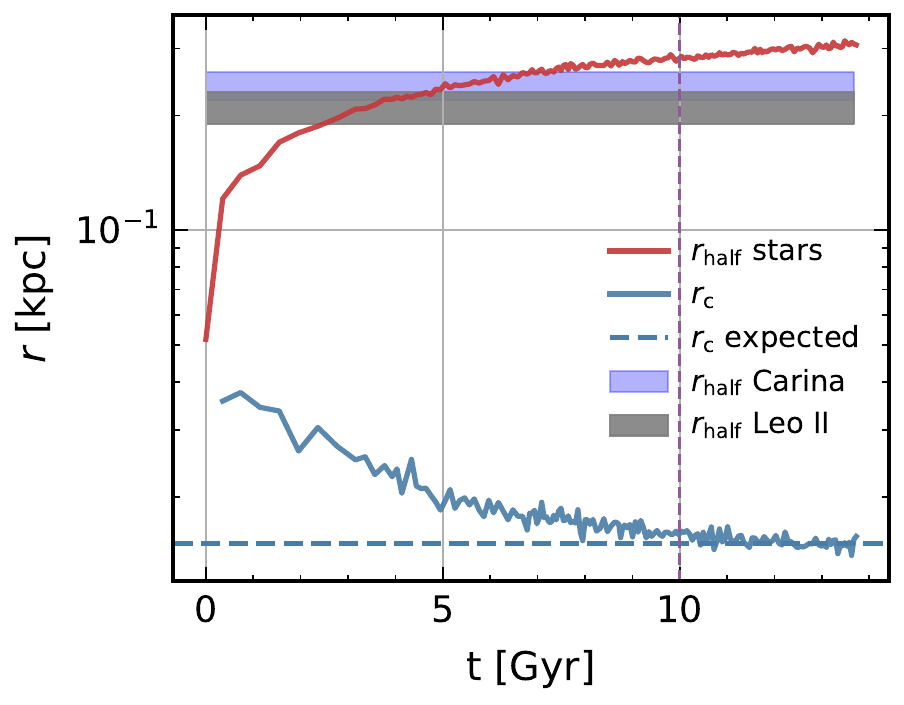}
    \caption{Results for LeoII/Carina for $m= 10^{-20}$ eV. Panels explanation is the same as upper panels explanation of Fig.~\ref{fig:NFW_5em21}.}
    \label{fig:LeoII_1e20}
\end{figure*}

Finally, we show results for $m = 10^{-20}$ eV for LeoII/Carina in Fig.~\ref{fig:LeoII_1e20}. Although dynamical heating is still noticeably present, it is not fast enough to put any significant tension as far as the half-light radius is concerned.

\section{\boldmath More simulations details} \label{s:sim_details}
We list the initial conditions of our simulation suite, together with the $r_{\rm half}$ reached after 10 Gyr and the value of $\sigma_{\rm los}$ on the innermost bin after 10 Gyr, in Tab.~\ref{tab:details}. 

\begin{figure*}
    \centering
\includegraphics[width=0.49\textwidth]{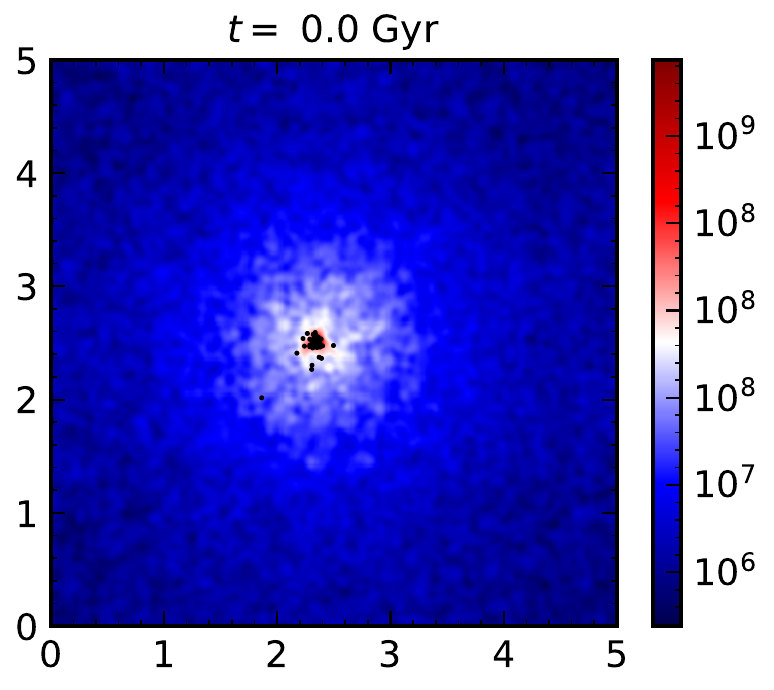}
\includegraphics[width=0.49\textwidth]{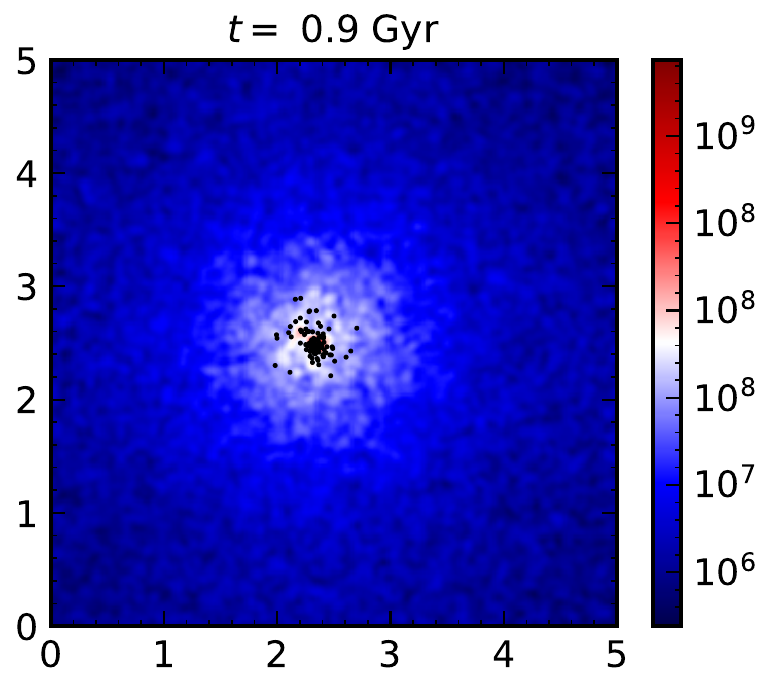}
\includegraphics[width=0.49\textwidth]{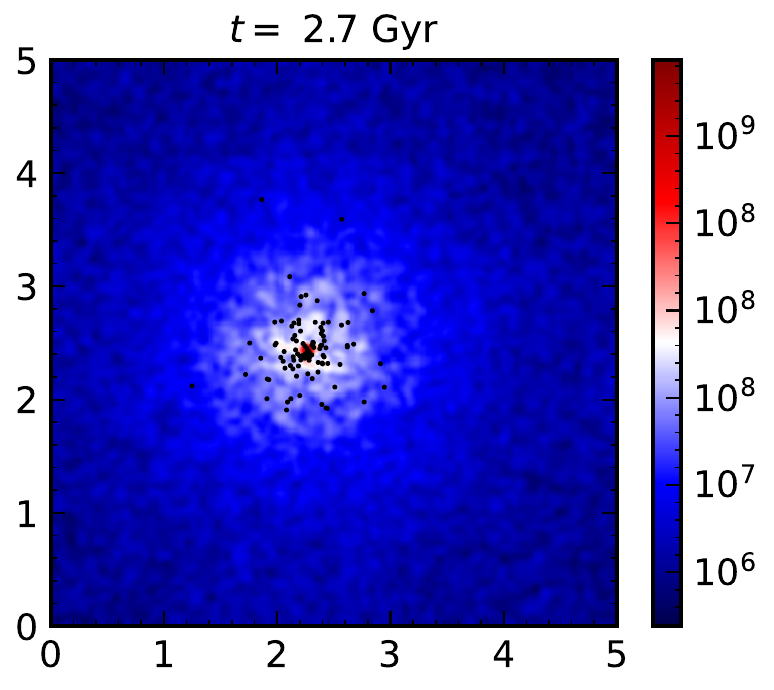}
\includegraphics[width=0.49\textwidth]{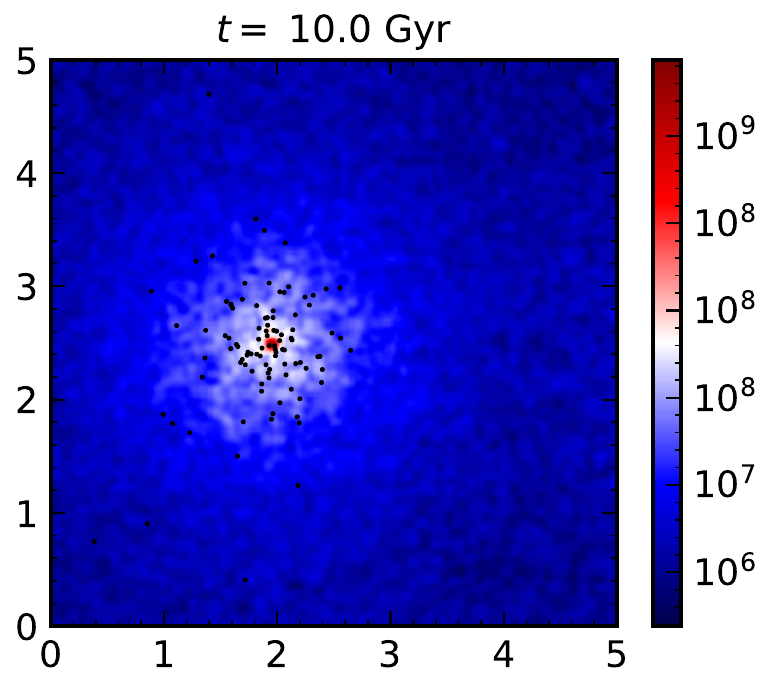}
\caption{Column density snapshots (in $M_\odot /$kpc$^2$) of the Leo II simulation with $m=\SI{5e-21}{\electronvolt}$ in a box of length $L = 5$ kpc, whose main results are shown in Fig.~\ref{fig:NFW_5em21_LeoII_main}. Black dots show 100 randomly selected stars (out of 10000 stars in the simulation).}
\label{fig:snaps}
\end{figure*}

We show column density snapshots of the Leo II simulation (for which the main results are shown in Fig.~\ref{fig:NFW_5em21_LeoII_main}), together with a sample of projected star positions, in Fig.~\ref{fig:snaps}. The growth of the half-light radius over time is apparent.

\begin{figure*}
    \centering
\includegraphics[width=0.7\linewidth]{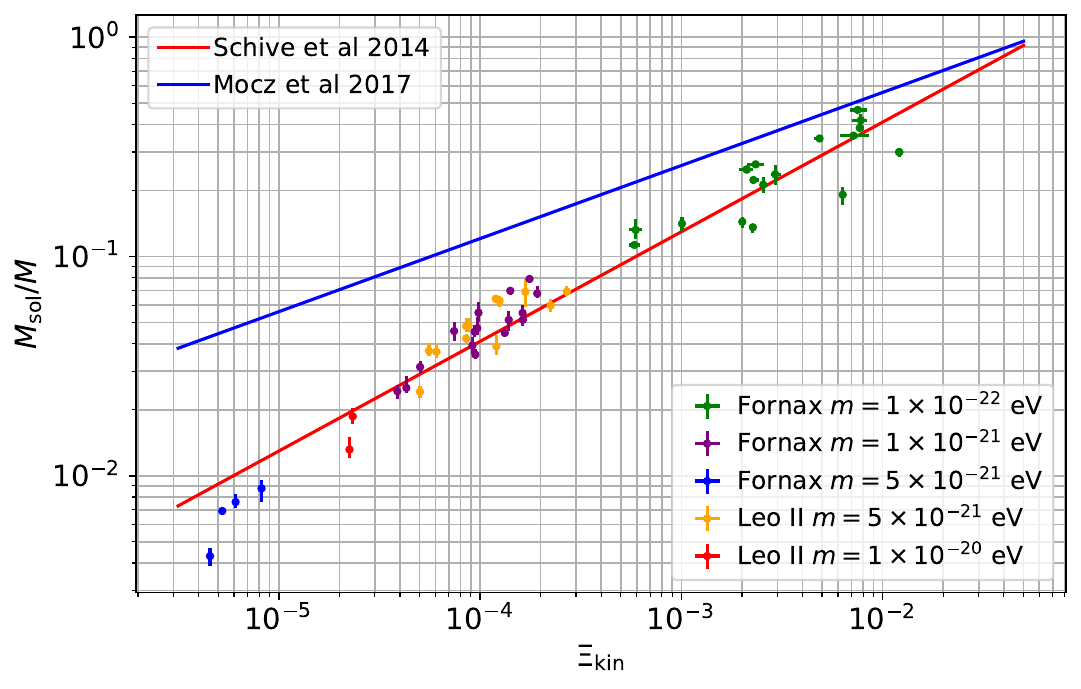}
    \caption{Soliton-halo relation points for our simulation suite. Different colored points refer to simulations with target galaxy either Fornax or LeoII/Carina and different particle masses. All of the simulations shown in the plot, give a reasonable description of the relevant galaxy's stellar LOSVD at large radius, outside the would-be soliton region. Blue line represents the relation of Ref.~\cite{Mocz:2017wlg} whereas the red line shows the relation of Ref.~\cite{Schive:2014hza}. As discussed in Ref.~\cite{Bar:2018acw,Blum:2025aaa}, the relation of~\cite{Mocz:2017wlg} should be understood as an upper limit.}
    \label{fig:sol_halo}
\end{figure*}

We show the core-halo diversity among our simulation suite in Fig.~\ref{fig:sol_halo}, where we show $M_{\rm sol}/M$ versus $\Xi_{\rm kin}$, where 
\begin{equation}
\Xi_{\rm kin} := \frac{E_{\rm kin}}{M^3} \frac{1}{G^2 m^2} \ .
\end{equation}
For comparison, in blue we show the soliton-halo relation discussed in Ref.~\cite{Mocz:2017wlg}, and in red the prediction from Ref.~\cite{Schive:2014hza}. 
Points are computed after $10 $ Gyr;
error bars show the $\pm 1$ Gyr spread around $1 $ Gyr of both $M_{\rm sol}/M$ and $\Xi_{\rm kin}$. As discussed in Ref.~\cite{Bar:2018acw,Blum:2025aaa}, the relation of~\cite{Mocz:2017wlg} should be understood as an upper limit.

\begin{table*}
\centering
\begin{tabular}{ccccccccccc}
\toprule
$m$ & Target & $\displaystyle\frac{L_{\rm box}}{{\rm kpc}}$ & Profile & $\displaystyle\frac{\rho_0}{M_\odot/{\rm pc}^3}$ & $\displaystyle\frac{r_{\rm s}}{{\rm kpc}}$ & $\displaystyle\frac{r^{\rm ini}_{\rm plum}}{{\rm kpc}}$ & $\beta$ & $\displaystyle\frac{r_{\rm half}^{10\, {\rm Gyr}}}{{\rm kpc}}$ & $\displaystyle\frac{\sigma^{10\, {\rm Gyr}}_{\rm los}}{{\rm km/s}}$ & Figure \\
\midrule
$ 1 \times 10^{-22}$ & Fornax & $40$ & NFW & $0.03$ & $1.5$ & $0.67$ & $0$ & $2.79^{+0.13}_{-0.08}$ & $20.5^{+1.7}_{-1.4}$ & \\
$ 1 \times 10^{-22}$ & Fornax & $40$ & Burkert & $0.05$ & $1.5$ & $0.67$ & $0$ & $2.76^{+0.00}_{-0.03}$ & $38.3^{+1.6}_{-3.8}$ & \\
$ 1 \times 10^{-22}$ & Fornax & $40$ & Burkert & $0.05$ & $1.5$ & $0.24$ & $0$ & $3.03^{+0.02}_{-0.16}$ & $35.8^{+1.2}_{-2.8}$ & \\
$ 1 \times 10^{-22}$ & Fornax & $40$ & NFW & $0.015$ & $1.5$ & $0.2$ & $0$ & $2.41^{+0.18}_{-0.25}$ & $16.7^{+3.4}_{-2.1}$ & \\
$ 1 \times 10^{-22}$ & Fornax & $40$ & NFW & $0.015$ & $1.5$ & $1.2$ & $0$ & $3.27^{+0.17}_{-0.03}$ & $14.8^{+2.3}_{-0.0}$ & \\
$ 1 \times 10^{-22}$ & Fornax & $40$ & NFW & $0.019$ & $1.5$ & $0.6$ & $0$ & $3.72^{+0.02}_{-0.52}$ & $10.5^{+2.6}_{-0.5}$ & \\
$ 1 \times 10^{-22}$ & Fornax & $40$ & NFW & $0.019$ & $1.5$ & $0.2$ & $0$ & $2.44^{+0.08}_{-0.07}$ & $22.4^{+2.2}_{-0.3}$ & \\
$ 1 \times 10^{-22}$ & Fornax & $40$ & Burkert & $0.016$ & $1.5$ & $0.6$ & $0$ & $3.83^{+0.01}_{-0.42}$ & $11.8^{+1.6}_{-1.4}$ & \\
$ 1 \times 10^{-22}$ & Fornax & $40$ & Burkert & $0.016$ & $1.5$ & $0.2$ & $0$ & $2.65^{+0.06}_{-0.05}$ & $23.6^{+0.2}_{-1.8}$ & \\
$ 1 \times 10^{-22}$ & Fornax & $40$ & NFW & $0.019$ & $1.5$ & $0.2$ & $0$ & $2.51^{+0.04}_{-0.10}$ & $24.7^{+0.3}_{-2.3}$ & \\
$ 1 \times 10^{-22}$ & Fornax & $40$ & NFW & $0.008$ & $1.5$ & $0.2$ & $0$ & $1.72^{+0.19}_{-0.21}$ & $17.4^{+3.4}_{-0.5}$ & \\
$ 1 \times 10^{-22}$ & Fornax & $40$ & NFW & $0.011$ & $1.5$ & $0.2$ & $0.0$ & $1.93^{+0.05}_{-0.04}$ & $23.9^{+1.9}_{-2.1}$ & \\
$ 1 \times 10^{-22}$ & Fornax & $40$ & Burkert & $0.008$ & $1.5$ & $0.2$ & $0.0$ & $1.79^{+0.32}_{-0.22}$ & $35.7^{+1.1}_{-1.1}$ & \\
$  1 \times 10^{-22}$ & Fornax & $40$ & Burkert & $0.0053$ & $1.5$ & $0.2$ & $0.0$ & $2.35^{+0.12}_{-0.28}$ & $16.3^{+0.4}_{-3.0}$ & Fig.~\ref{fig:1em22_large_main},~\ref{fig:1em22_large} \\
$ 1 \times 10^{-22}$ & Fornax & $40$ & NFW & $0.008$ & $1.5$ & $0.4$ & $0.0$ & $3.41^{+0.86}_{-0.21}$ & $5.8^{+1.0}_{-1.1}$ & \\
$ 1 \times 10^{-22}$ & Fornax & $40$ & Burkert & $0.004$ & $1.5$ & $0.2$ & $0.0$ & $1.61^{+0.03}_{-0.55}$ & $0.8^{+1.1}_{-0.1}$ & \\
$ 1 \times 10^{-22}$ & Fornax & $40$ & NFW & $0.008$ & $1.5$ & $0.2$ & $0.0$ & $1.60^{+0.20}_{-0.10}$ & $15.8^{+0.7}_{-2.2}$ & \\
$ 1 \times 10^{-22}$ & Fornax & $40$ & Burkert & $0.004$ & $1.5$ & $0.2$ & $0.0$ & $1.40^{+0.37}_{-0.06}$ & $0.8^{+1.0}_{-0.1}$ & \\
$ 1 \times 10^{-22}$ & Fornax & $40$ & Burkert & $0.016$ & $1.5$ & $0.4$ & $0.0$ & $1.63^{+0.10}_{-0.22}$ & $16.6^{+1.5}_{-0.7}$ & \\
$ 5 \times 10^{-22}$ & Fornax & $25$ & NFW & $0.0029$ & $3.5$ & $0.38$ & $-0.3$ & $1.68^{+0.04}_{-0.07}$ & $12.8^{+1.9}_{-0.6}$ & \\
$ 5 \times 10^{-22}$ & Fornax & $25$ & NFW & $0.0077$ & $1.5$ & $0.2$ & $-0.3$ & $1.91^{+0.07}_{-0.12}$ & $9.2^{+0.5}_{-0.2}$ & \\
$ 1 \times 10^{-21}$ & Fornax & $12$ & NFW & $0.03$ & $1.5$ & $0.76$ & $0$ & $1.11^{+0.03}_{-0.02}$ & $18.1^{+5.4}_{-0.3}$ & \\
$ 1 \times 10^{-21}$ & Fornax & $12$ & NFW & $0.012$ & $1.5$ & $0.24$ & $0$ & $0.90^{+0.02}_{-0.04}$ & $20.3^{+1.1}_{-3.2}$ & Fig.~\ref{fig:NFW_1em21Beta0ini} \\
$ 1 \times 10^{-21}$ & Fornax & $12$ & NFW & $0.012$ & $1.5$ & $1.4$ & $0$ & $1.43^{+0.00}_{-0.05}$ & $12.3^{+0.6}_{-0.6}$ & \\
$ 1 \times 10^{-21}$ & Fornax & $12$ & NFW & $0.0099$ & $1.5$ & $0.76$ & $0$ & $1.18^{+0.05}_{-0.03}$ & $11.4^{+1.2}_{-0.9}$ & \\
$ 1 \times 10^{-21}$ & Fornax & $12$ & NFW & $0.0099$ & $1.5$ & $0.24$ & $0$ & $1.08^{+0.06}_{-0.07}$ & $12.4^{+1.6}_{-0.9}$ & \\
$  1 \times 10^{-21}$ & Fornax & $12$ & NFW & $0.016$ & $1.5$ & $0.76$ & $0$ & $1.24^{+0.02}_{-0.04}$ & $14.3^{+0.2}_{-3.0}$ & Fig.~\ref{fig:NFW_1em21_main}\\
$  1 \times 10^{-21}$ & Fornax & $12$ & NFW & $0.016$ & $1.5$ & $0.24$ & $0$ & $0.94^{+0.03}_{-0.03}$ & $15.6^{+3.1}_{-1.0}$ & Fig.~\ref{fig:NFW_1em21_main},~\ref{fig:init_density},~\ref{fig:Jeans_time}  \\
$  1 \times 10^{-21}$ & Fornax & $12$ & NFW & $0.016$ & $1.5$ & $0.24$ & $-0.3$ & $0.89^{+0.04}_{-0.04}$ & $18.8^{+2.6}_{-1.2}$ & Fig.~\ref{fig:NFW_1em21Beta} \\
$  1 \times 10^{-21}$ & Fornax & $12$ & NFW & $0.016$ & $1.5$ & $0.76$ & $-0.3$ & $1.21^{+0.01}_{-0.06}$ & $15.0^{+1.8}_{-2.8}$ & \\
$ 1 \times 10^{-21}$ & Fornax & $12$ & NFW & $0.016$ & $1.5$ & $0.24$ & $-0.3$ & $1.01^{+0.05}_{-0.03}$ & $15.7^{+3.7}_{-0.2}$ & \\
$ 1 \times 10^{-21}$ & Fornax & $10$ & NFW & $0.031$ & $1.0$ & $0.2$ & $-0.2$ & $0.90^{+0.06}_{-0.03}$ & $16.9^{+1.6}_{-2.6}$ & \\
$ 1 \times 10^{-21}$ & Fornax & $12$ & Burkert & $0.03$ & $1.5$ & $0.24$ & $0.0$ & $0.85^{+0.04}_{-0.08}$ & $18.8^{+1.9}_{-1.5}$ & \\
$ 1 \times 10^{-21}$ & Fornax & $12$ & Burkert & $0.033$ & $1.5$ & $0.48$ & $0.0$ & $0.96^{+0.03}_{-0.04}$ & $20.7^{+0.3}_{-2.8}$ & \\
$ 1 \times 10^{-21}$ & Fornax & $12$ & NFW & $0.017$ & $1.2$ & $0.24$ & $-0.2$ & $0.78^{+0.00}_{-0.01}$ & $20.5^{+0.1}_{-0.3}$ & \\
$ 1 \times 10^{-21}$ & Fornax & $10$ & NFW & $0.014$ & $1.6$ & $0.2$ & $-0.2$ & $0.92^{+0.05}_{-0.01}$ & $15.2^{+2.1}_{-2.2}$ & \\
$ 1 \times 10^{-21}$ & Fornax & $12$ & Burkert & $0.046$ & $1.0$ & $0.48$ & $0.0$ & $0.92^{+0.03}_{-0.06}$ & $20.8^{+2.5}_{-2.4}$ & \\
$ 5 \times 10^{-21}$ & Fornax* & $8$ & NFW & $0.027$ & $1.5$ & $0.72$ & $0$ & $0.70^{+0.01}_{-0.01}$ & $13.7^{+1.0}_{-1.0}$ & \\
$ 5 \times 10^{-21}$ & Fornax* & $8$ & NFW & $0.015$ & $1.5$ & $0.76$ & $-0.2$ & $0.73^{+0.00}_{-0.02}$ & $11.8^{+0.7}_{-2.0}$ & \\
$ 5 \times 10^{-21}$ & Fornax* & $8$ & NFW & $0.023$ & $1.5$ & $0.76$ & $0.0$ & $0.69^{+0.00}_{-0.00}$ & $13.7^{+0.6}_{-0.3}$ &  \\
$  5 \times 10^{-21}$ & Fornax* & $8$ & NFW & $0.02$ & $1.5$ & $0.76$ & $0.0$ & $0.71^{+0.01}_{-0.02}$ & $12.8^{+0.9}_{-1.3}$ & Fig.~\ref{fig:NFW_5em21} \\
$  5 \times 10^{-21}$ & Leo II/Carina & $5$ & NFW & $0.018$ & $0.7$ & $0.075$ & $-0.3$ & $0.43^{+0.03}_{-0.01}$ & $7.5^{+1.6}_{-0.1}$ & Fig.~\ref{fig:NFW_5em21_LeoII_main} \\
$  5 \times 10^{-21}$ & Leo II/Carina & $5$ & NFW & $0.018$ & $0.7$ & $0.025$ & $-0.3$ & $0.38^{+0.01}_{-0.02}$ & $8.9^{+1.6}_{-0.5}$ &  Fig.~\ref{fig:NFW_5em21_LeoII_main},~\ref{fig:snaps}\\
$ 5 \times 10^{-21}$ & Leo II/Carina & $5$ & NFW & $0.01$ & $0.8$ & $0.05$ & $0.0$ & $0.47^{+0.00}_{-0.03}$ & $5.8^{+0.8}_{-0.1}$ & \\
$ 5 \times 10^{-21}$ & Leo II/Carina & $5$ & NFW & $0.039$ & $0.4$ & $0.05$ & $0.0$ & $0.42^{+0.03}_{-0.01}$ & $9.2^{+0.0}_{-2.0}$ & \\
$ 5 \times 10^{-21}$ & Leo II/Carina & $5$ & NFW & $0.03$ & $0.5$ & $0.05$ & $0.0$ & $0.39^{+0.02}_{-0.01}$ & $9.7^{+0.2}_{-2.1}$ & \\
$ 5 \times 10^{-21}$ & Leo II/Carina & $5$ & NFW & $0.027$ & $0.42$ & $0.1$ & $-0.2$ & $0.49^{+0.01}_{-0.03}$ & $5.1^{+0.3}_{-0.7}$ & \\
$  5 \times 10^{-21}$ & Leo II/Carina* & $5$ & NFW & $0.018$ & $0.7$ & $0.025$ & $-0.3$ & $0.40^{+0.00}_{-0.02}$ & $9.1^{+1.1}_{-0.6}$ & Fig.~\ref{fig:LeoII_highres}\\
$ 5 \times 10^{-21}$ & Leo II/Carina & $5$ & Burkert & $0.02$ & $0.8$ & $0.1$ & $0.0$ & $0.38^{+0.01}_{-0.03}$ & $7.3^{+0.8}_{-0.1}$ & \\
$ 5 \times 10^{-21}$ & Leo II/Carina & $5$ & Burkert & $0.048$ & $0.5$ & $0.1$ & $0.0$ & $0.38^{+0.01}_{-0.02}$ & $9.0^{+1.4}_{-0.8}$ & \\
$ 5 \times 10^{-21}$ & Leo II/Carina & $5$ & Burkert & $0.052$ & $0.5$ & $0.1$ & $0.0$ & $0.36^{+0.02}_{-0.02}$ & $10.5^{+0.9}_{-2.1}$ & \\
$ 5 \times 10^{-21}$ & Leo II/Carina & $5$ & Burkert & $0.096$ & $0.3$ & $0.05$ & $0.0$ & $0.34^{+0.01}_{-0.02}$ & $9.5^{+1.2}_{-1.1}$ & \\
$  1 \times 10^{-20}$ & Leo II/Carina* & $4$ & NFW & $0.025$ & $0.7$ & $0.1$ & $0.0$ & $0.29^{+0.00}_{-0.01}$ & $6.8^{+0.7}_{-0.5}$ & Fig.~\ref{fig:LeoII_1e20} \\
$ 1 \times 10^{-20}$ & Leo II/Carina* & $4$ & NFW & $0.025$ & $0.7$ & $0.052$ & $0.0$ & $0.28^{+0.00}_{-0.02}$ & $7.9^{+1.4}_{-0.8}$ & \\
\bottomrule
\end{tabular}
\caption{Overview of our simulations suite. {\bf Column 1}: particle mass in eV; {\bf column 2}: system target of the simulation, asterisk (*) denotes an high resolution simulation ($512^3$ points compared with the usual $256^3$); {\bf column 3}: length of the box in kpc; {\bf column 4}: profile used for the initialization; {\bf column 5}: $\rho_0$ parameter in $M_\odot/{\rm pc}^3$; {\bf column 6}: $r_{\rm s} $ in kpc; {\bf column 7}: initial $r_{\rm plum} $ for the stars distribution in kpc; {\bf column 8}: initial $\beta$ anisotropy parameter; {\bf column 9}: $r_{\rm half}$ in kpc after $10$ Gyr, error bars show the spread for $10\pm 1$ Gyr; {\bf column 10}: $\sigma_{\rm los}$ after $10$ Gyr in the innermost bin, error bars show the spread for $10\pm 1$ Gyr; {\bf column 11}: figures in which the simulation results appear. All of the simulations in the table, are tailored to give a reasonable description of the relevant galaxy's stellar LOSVD at large radius, outside the would-be soliton region.}
\label{tab:details}
\end{table*}

\section{Tidally stripped single-core halos} \label{s:tidal}

Refs.~\cite{Chan:2025hhg,Yang:2025bae} pointed out that tidal stripping of the ULDM halo may significantly reduce the dynamical heating for $m \sim \SI{1e-22}{\electronvolt}$. We argue here that this is unlikely to affect our results for $m \gtrsim \SI{5e-22}{\electronvolt}$.

\begin{figure}
    \centering
    \includegraphics[width=0.99\linewidth]{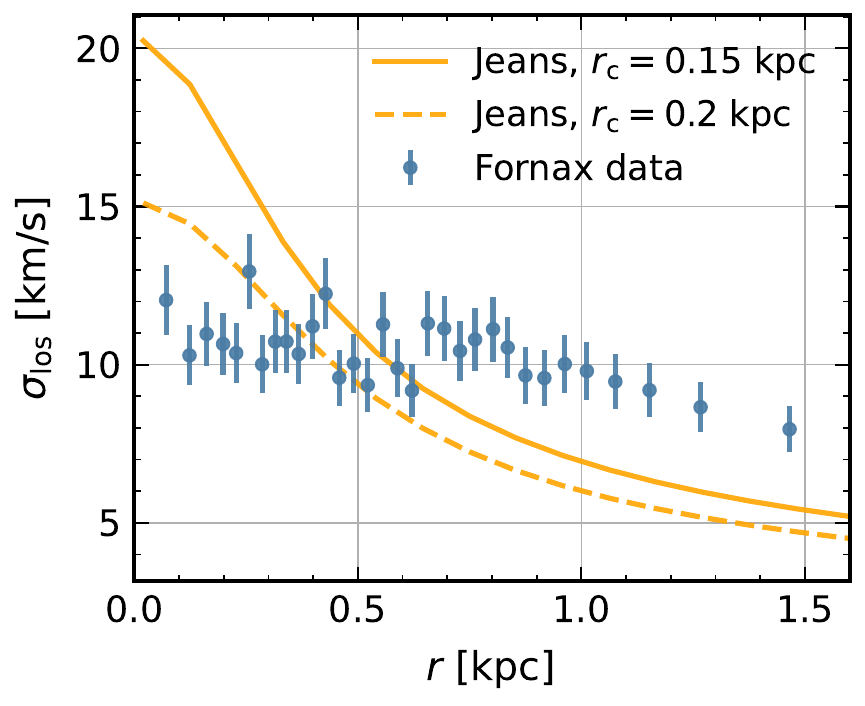}
    \includegraphics[width=0.99\linewidth]{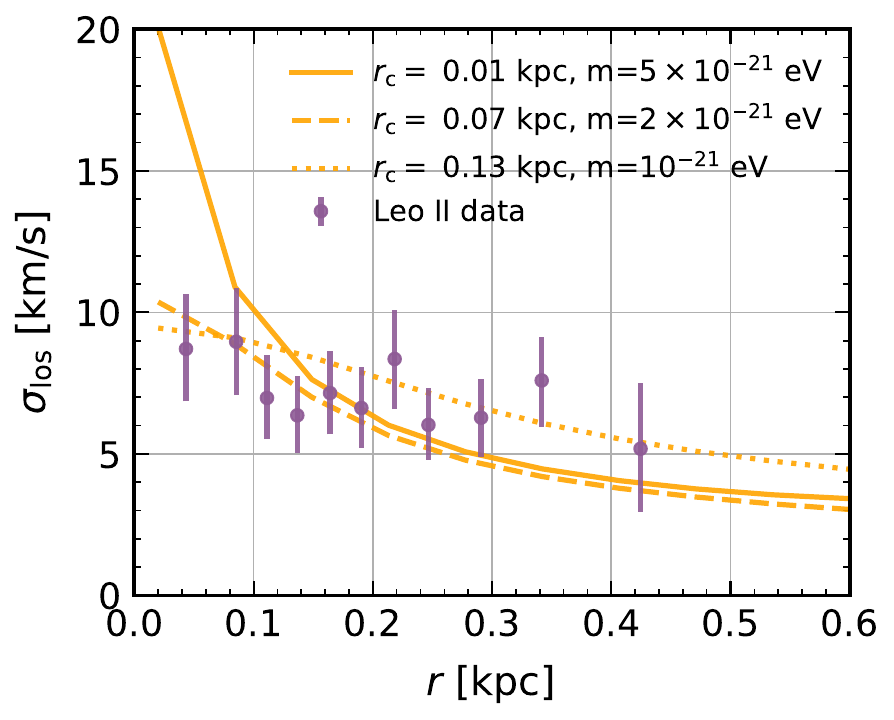}
    \caption{ \textbf{Top}: Trying to fit Fornax data with a single soliton with ULDM mass $m=\SI{5e-22}{\electronvolt}$, via simple Jeans analysis. We present two different core radii, to illustrate the difficulty in fitting both tail and inner LOSVD points, similar to what happens for $m=\SI{1e-21}{\electronvolt}$. \textbf{Bottom}: the same but for Leo II, and ULDM mass $m=\SI{5e-21}{\electronvolt}$ (solid), $m=\SI{2e-21}{\electronvolt}$ (dashed), $m=\SI{e-21}{\electronvolt}$ (dotted). }
    \label{fig:jeans_single_sol}
\end{figure}

First, a tidally stripped halo means that in practice a bare soliton should fit the full LOSVD. This is not possible for Fornax for $m\gtrsim \SI{5e-22}{\electronvolt}$, as a simple Jeans analysis can show, see Fig.~\ref{fig:jeans_single_sol} top. 

For the other dwarfs we considered, an estimate of the tidal radius for Leo~II yields
\begin{equation}
r_{\rm t} \sim 6 \, \mathrm{kpc} \left(\frac{D}{233 \, \mathrm{kpc}}\right)
\left(\frac{10^{12}  M_\odot}{M_{\rm MW}}\right)^{1/3}
\left(\frac{M_{\rm Leo  II}}{10^7  M_\odot}\right)^{1/3},
\end{equation}
which is about 30 times larger than the stellar half-light radius and also much larger than the soliton core for all masses of interest. Even if the pericenter of Leo II's orbit were a factor of a few smaller than its distance today, there would still be a sizable separation of scales between $r_{\rm t}$ and $r_{\text{half}},\,r_{\rm c}$, suggesting that tidal stripping is likely not significant. Indeed, SDSS I data for Leo II did not reveal any significant signs of tidal distortion~\cite{2007AJ....134.1938C}\footnote{The absence of obvious morphological disturbances (e.g.\ isophotal twisting, elongation, or extra-tidal features) is not, by itself, a robust indicator of negligible tidal evolution: such signatures can be transient, are most prominent near/after pericenter, and may only persist for a few internal dynamical times, which are indeed much shorter than typical orbital periods. However, the fact that SDSS~I did not find compelling stellar tidal features in Leo~II~\cite{2007AJ....134.1938C} and the fact that Leo~II is very spherical, are good consistency checks.} Notice that the most relevant quantity for dynamical heating is how many energy eigenstates are removed by tidal stripping~\cite{Eberhardt:2025lbx, Yang:2025bae, May:2025ppj}. Only when a large fraction of the excited eigenstates is stripped does the heating become significantly reduced (see, e.g., Fig.~12 of Ref.~\cite{Eberhardt:2025lbx}). Therefore, the relevant diagnostic is the ratio between the soliton size (which corresponds roughly to the lowest-energy eigenstate) and the tidal radius.

We also note that the kinematics data disfavors a bare soliton fit of Leo~II for \(m \sim 5 \times 10^{-21}\,\mathrm{eV}\), as shown by the solid line in Fig.~\ref{fig:jeans_single_sol} bottom, where we compare, as an example, Leo~II data with a bare soliton model. Not only is it impossible to obtain a good fit to the data (even with extreme value of the anisotropy parameter), but, for simple scaling reasons, the core radius is forced to become extremely small -- about 20 times smaller than the stellar half-light radius -- making it unlikely that the halo could have been stripped away without also stripping the stellar population at larger distances. The fit starts to improve for $2\, \times \, 10^{-21}$ eV (dashed line in Fig.19 bottom), but the core radius remains much smaller than the half light radius of the system. For $m \lesssim 10^{-21}$ eV the ULDM core is large enough to provide a good fit to the data with a core radius comparable to the half light radius of the system (dotted line in Fig.~19 bottom).

In conclusion, the ULDM mass range $ 5\times10^{-22} \lesssim m/\text{eV} \lesssim 5\times10^{-21} $ seems disfavored even when tidal stripping is taken into account. We reserve direct verification implementing the Milky Way halo tidal field in several dwarf satellite galaxies to future work.

\bibliographystyle{bibi}
\bibliography{biblio}
\end{document}